\documentclass[twocolumn,numberedappendix, twocolappendix, apj]{openjournal}
\usepackage{newtxtext,newtxmath}
\usepackage{amsmath}
\usepackage{amssymb}
\usepackage{graphicx}
\usepackage{dcolumn}
\usepackage{bm}
\usepackage[dvipsnames]{xcolor}
\usepackage{xspace}
\usepackage[colorlinks=true, allcolors=blue]{hyperref}
\usepackage{ulem}
\usepackage{lipsum}
\newcommand{\desyo}{DES-Y1\xspace}
\newcommand{\goodpar}{\vec{\Omega}}
\newcommand{\badpar}{\vec{n}}
\newcommand{\nv}{\hat{\bf n}}

\newcommand{\Om}{\Omega_{\rm m}}
\newcommand{\sig}{\sigma_{\rm 8}}

\newcommand{\Ob}{\Omega_{\rm b}}
\newcommand{\cN}{\mathcal{N}}
\newcommand{\ns}{n_{\rm s}}
\newcommand{\redmagic}{\textsc{redMaGiC}\xspace}
\newcommand{\mcal}{\textsc{Metacalibration}\xspace}

\begin{document}

\title{Cosmology with 6 parameters in the Stage-IV era: efficient marginalisation over nuisance parameters}

\author{B. Hadzhyiska$^{1,2,*}$}
\author{K. Wolz$^{3,4}$}
\author{S. Azzoni$^{5,6}$}
\author{D. Alonso$^{5}$}
\author{C. Garc{\'\i}a-Garc{\'\i}a$^{5}$}
\author{J. Ruiz-Zapatero$^{5}$}
\author{A. Slosar$^{7}$}
\email{$^*$boryanah@berkeley.edu}
\affiliation{$^1$Miller Institute for Basic Research in Science, University of California, Berkeley, CA, 94720, USA}
\affiliation{$^2$Physics Division, Lawrence Berkeley National Laboratory, Berkeley, CA 94720}
\affiliation{$^3$International School for Advanced Studies (SISSA), Via Bonomea 265, 34136 Trieste, Italy}
\affiliation{$^4$National Institute for Nuclear Physics (INFN) -- Sezione di Trieste, Via Valerio 2, 34127 Trieste, Italy}
\affiliation{$^5$Department of Physics, University of Oxford, Denys Wilkinson Building, Keble Road, Oxford OX1 3RH, United Kingdom}
\affiliation{$^6$Kavli Institute for the Physics and Mathematics of the Universe (Kavli IPMU, WPI), UTIAS, The University of Tokyo, Kashiwa, Chiba 277-8583, Japan}
\affiliation{$^7$Physics Department, Brookhaven National Laboratory, Upton NY 11973, USA}
\date{\today}

\begin{abstract}
  The analysis of photometric large-scale structure data is often complicated by the need to account for many observational and astrophysical systematics. The elaborate models needed to describe them often introduce many ``nuisance parameters'', which can be a major inhibitor of an efficient parameter inference. In this paper, we introduce an approximate method to analytically marginalise over a large number of nuisance parameters based on the  Laplace approximation. We discuss the mathematics of the method, its relation to concepts such as volume effects and profile likelihood, and show that it can be further simplified for calibratable systematics by linearising the dependence of the theory on the associated parameters. We quantify the accuracy of this approach by comparing it with traditional sampling methods in the context of existing data from the Dark Energy Survey, as well as futuristic Stage-IV photometric data. The linearised version of the method is able to obtain parameter constraints that are virtually equivalent to those found by exploring the full parameter space for a large number of calibratable nuisance parameters, while reducing the computation time by a factor 3-10. Furthermore, the non-linearised approach is able to analytically marginalise over a large number of parameters, returning constraints that are virtually indistinguishable from the brute-force method in most cases, accurately reproducing both the marginalised uncertainty on cosmological parameters, and the impact of volume effects associated with this marginalisation. We provide simple recipes to diagnose when the approximations made by the method fail, and one should thus resort to traditional methods. The gains in sampling efficiency associated with this method enable the joint analysis of multiple surveys, typically hindered by the large number of nuisance parameters needed to describe them.
 \end{abstract}
\maketitle

\section{Introduction}\label{sec:intro}
  The $\Lambda$ Cold Dark Matter ($\Lambda$CDM) model of cosmology offers a compelling explanation for a wide variety of observations despite the fact that the nature of its dominant components, dark matter and dark energy, remains unknown. As the statistical power of cosmological experiments grows in the next decade, our ability to stress test the $\Lambda$CDM model of the Universe will improve immensely and provide powerful constraints on its dark ingredients. In addition, these new experiments may shed light over discrepancies between early- and late-Universe-derived cosmological constraints that have recently emerged. 

  Of highest relevance to this study is the tension affecting the measurement of the amplitude of matter fluctuations, $S_8 \equiv \sigma_8 (\Omega_m/0.3)^{0.5}$, where $\sigma_8$ is the root mean square of matter fluctuations on an 8 ${\rm Mpc}/h$ scale, and $\Omega_m$ is the fractional energy density in non-relativistic matter. In particular, the latest prediction from \textit{Planck} CMB data finds its value \citep{2020A&A...641A...6P} to be $\sim$2-3$\sigma$ higher than the analogous measurement from photometric surveys such as KiDS-1000 (KiDS), the Dark Energy Survey (DES) and Hyper Suprime-Cam (HSC) \citep{2021A&A...646A.140H,2022PhRvD.105b3520A,
  2018PhRvD..98d3526A,2015MNRAS.451.2877M,10.1093/pasj/psz138
  }.  As data from current and future cosmological surveys such as the Legacy Survey of Space and Time (LSST), at the Vera Rubin Observatory \citep{2012arXiv1211.0310L}, the Nancy Grace Roman Space Telescope \citep{2015arXiv150303757S}, or the Euclid satellite \citep{2018LRR....21....2A} starts to trickle in, it is of paramount importance for cosmologists to devise robust tests of their analysis pipelines, and to construct rigorous theoretical frameworks to better understand these tensions and extract maximal cosmological information. 

  Photometric surveys offer a pathway to resolving many issues of the standard paradigm by providing measurements of the clustering and weak gravitational lensing of millions and soon billions of galaxies on the sky. In particular, the so-called ``3$\times$2pt'' analysis (the joint analysis of galaxy clustering and cosmic shear in tomographic bins) offers a powerful tool to cosmologists with the potential to break degeneracies between cosmological and astrophysical parameters and yields stringent constraints \citep{2021A&A...646A.140H}. Such tomographic analyses have a leverage on the $S_8$ and $\Omega_m$ tensions and can yield $\sim$1-10\%-level constraints \citep[e.g.,][]{2020JCAP...03..044N,2021JCAP...09..020H, 2105.12108}. Nonetheless, it is still unclear whether this tension is driven by systematic and modeling errors in the analysis or by new physics.

  Large-scale structure surveys in general are affected by a large number of observational systematic uncertainties, as well as uncertainties in the physical relation between the astrophysical systems that form the basis of their observables, and the underlying cosmological quantities we wish to constrain (commonly labelled ``astrophysical systematics''). With increasing statistical power, new sources of systematic uncertainty become more relevant, and previously known systematics require more detailed models. This inevitably leads to an inflation in the number of nuisance parameters relative to the number of the physically meaningful cosmological parameters. Examples of these systematics are: uncertainties in the redshift distribution of the samples, errors in the measured galaxy shapes (relevant for weak lensing studies), the link between the abundance of galaxies and the underlying matter overdensities, and the intrinsic alignments between galaxy orientations and the local structures. Moreover, since the future of cosmology lies in the combination of multiple observational probes, the simultaneous modeling of these effects across different datasets, and the efficient sampling of the resulting (typically large) parameter space is bound to become a top priority in the next few years.

  To illustrate this, the latest analysis of galaxy clustering and weak gravitational lensing carried out by the Dark Energy Survey (DES) \citep{2022PhRvD.105b3520A} included 6 cosmological parameters ($\Lambda$CDM and the total neutrino mass) and 25 nuisance parameters. Due to the curse of dimensionality, this increase in the number of parameters leads to strong inefficiencies in the standard rejection sampling algorithms used to explore the parameter space. The resulting parameter chains take long times to converge (e.g. days or weeks for state-of-the-art datasets), even though we are ultimately only interested in the marginal posterior distribution of a much smaller parameter space.

  This problem can be addressed through various approaches. On one hand, the use of gradient-based sampling algorithms, such as Hamiltonian Monte-Carlo approaches \citep{HMC, NUTS}, or other methods that are well-suited to multiple-parameter problems, can significantly speed up the sampler convergence time and provide constraints in a reasonable amount of time. Recently, \citet{RZP} validated the analytical marginalisation of redshift calibration models with up to $\sim100$ parameters using self-tuning Hamiltonian Monte-Carlo (a problem that has also been addressed by e.g. \cite{2021A&A...650A.148S,2023MNRAS.518..709Z} using other methods). However, the performance of the marginalisation method depends on the choice of nuisance parameters, on one hand through their effect on the theory prediction, and on the other hand through their priors.

  For Gaussian data with parameters affecting the theory predictions to linear order, this can be effectively done by modifying the covariance matrix and fixing the marginalised parameters \citep{1992ApJ...398..169R}. For more complex parameters that appear as higher-order terms in the model such as the galaxy bias parameters, exact analytic marginalisation is not generally possible. One may resort to Gibbs sampling-like schemes \citep{gibbs}, where this marginalisation is done numerically on the fly, and which can potentially lead to significant speed-ups in the sampling process. However, the efficiency of this approach depends on the effective acceptance rate of the marginalisation step, and on the degeneracy between nuisance and cosmological parameters. Here, we propose a general technique that allows for approximate analytical marginalisation over both linear and non-linear nuisance parameters in an efficient manner. Similar methods have been put forward in the past \citep[e.g.,][]{2010MNRAS.408..865T}, with a variety of applications in mind \citep{2002MNRAS.335.1193B,2021A&A...650A.148S}. Here, we will quantify the validity of this method in the context of photometric 3$\times$2pt analyses with current data from DES, and futuristic Stage-IV-like data mimicking experiments such as LSST.

  This paper is organised as follows. In Section~\ref{sec:like}, we provide a general introduction to our analytical marginalisation method and discuss the interpretation of the various terms we define and their relevance to volume effects and parameter priors. In Section~\ref{sec:tomo}, we introduce relevant aspects of the analysis of cosmological photometric surveys and explore possible ways in which our method can be used to marginalise over linearisable and non-linearisable nuisance parameters in the model. In Section~\ref{sec:res}, we showcase the effect of employing our method both in terms of deriving unbiased constraints on the cosmological parameters and also in terms of the convergence time and performance. We summarise our findings and comment on their implications for the future of photometric survey analysis in Section~\ref{sec:conc}.

\section{Likelihoods and nuisance parameters}\label{sec:like}

  Our aim is to explore the posterior distribution $p(\vec{\theta}|{\bf d})$, where $\vec{\theta}$ is a set of model parameters, and ${\bf d}$ is the data. Using Bayes' theorem, the posterior can be written in terms of the likelihood $p({\bf d}|\vec{\theta})$ and prior $p(\vec{\theta})$ as
  \begin{equation}
    p(\vec{\theta}|{\bf d})\propto p({\bf d}|\vec{\theta})p(\vec{\theta}).
  \end{equation}
  We will decompose the model parameters as $\vec{\theta}=\{\goodpar,\badpar\}$, where $\goodpar$ is a set of parameters we care about (e.g. fundamental cosmological parameters), and $\badpar$ is a vector of nuisance parameters, describing various observational or theoretical uncertainties, which are largely irrelevant to the fundamental question being explored. In other words, the distribution we aim to obtain is the marginalised posterior
  \begin{equation}\label{eq:marg}
    p(\goodpar|{\bf d})=\int d\badpar\,p(\goodpar,\badpar|{\bf d}).
  \end{equation}

  For simplicity, in what follows, for any probability distribution $p(\vec{\theta})$, we will define the ``chi-squared'', $\chi^2$, as
  \begin{equation}
    \chi^2(\vec{\theta})\equiv-2\log p(\vec{\theta})+K,
  \end{equation}
  where $K$ is an arbitrary constant that does not depend on the random variables $\vec{\theta}$. Note that it is more common to define $\chi^2$ in terms of the likelihood, but in this paper, we will always apply it to the posterior, generalising to the presence of priors.
  \begin{figure*}
    \begin{center}
      \includegraphics[width=0.9\textwidth]{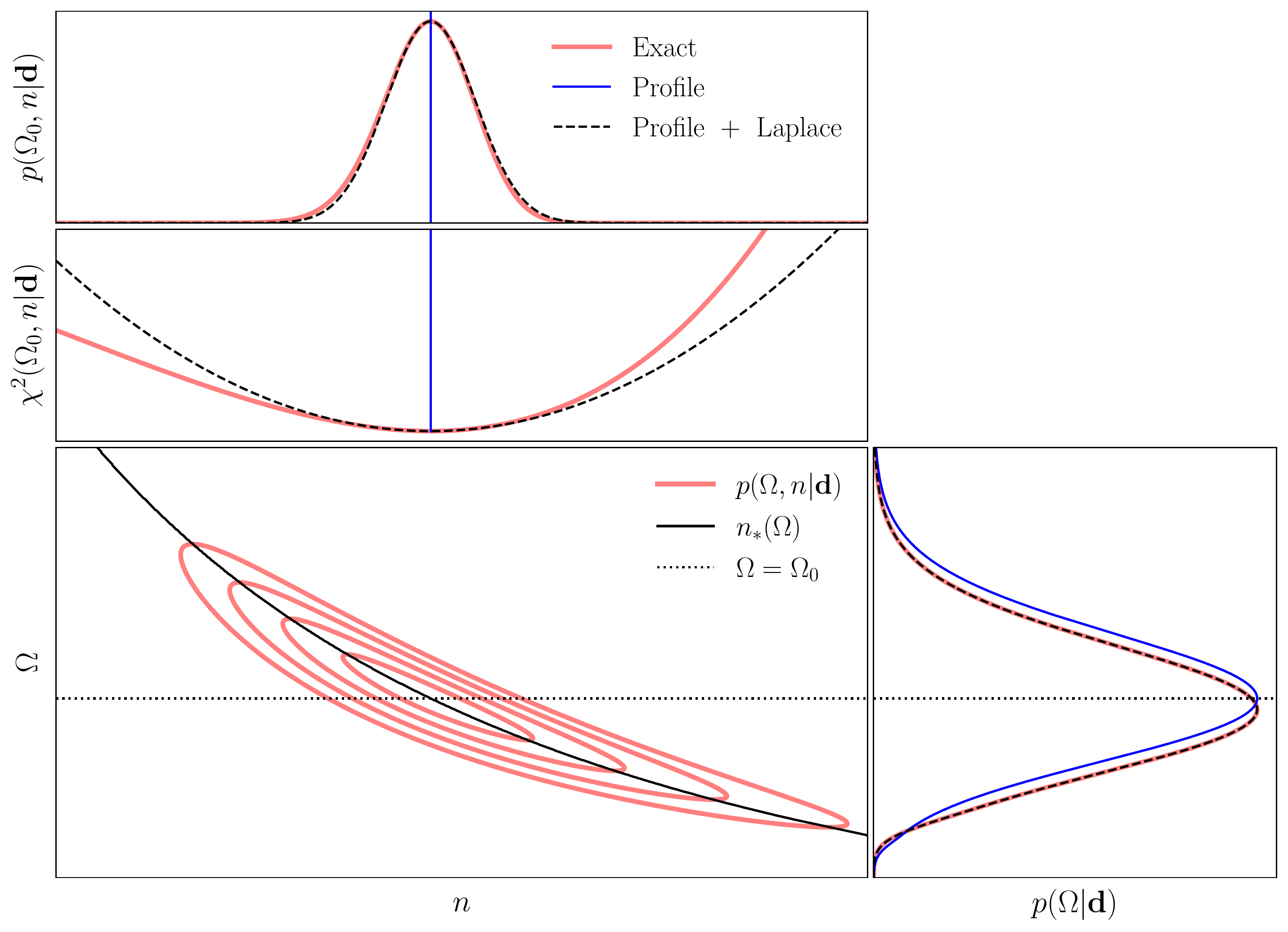}
      \caption{Joint posterior distribution on two parameters, $\Omega$ and $n$, with an approximate degeneracy of the form $n\Omega^{1.2}\sim{\rm const}$. The large bottom left panel shows the joint distribution as red contours, with the position of the best-fit value of $n$ as a function of $\Omega$ in solid black. The central panel shows the $\chi^2$ for a fixed value of $\Omega=\Omega_0$ (shown as the dotted line in the first panel). The red line shows the true $\chi^2$, while the dashed black line shows the Laplace approximation. The top panel shows the probability distribution along $\Omega=\Omega_0$ (given by $p\propto \exp(-\chi^2/2)$), with the exact distribution and its Laplace approximation in red and dashed black respectively. The bottom right panel shows the distribution of $\Omega$ marginalised over $n$. The exact result is shown in red, with its Laplace approximation shown in dashed black. The blue line shows the marginalised profile likelihood obtained by simply maximising the joint likelihood over $n$ for each $\Omega$. The Laplace approximation provides an excellent description of the marginalised distribution, while the profile likelihood returns a distribution with very similar width but centered, by construction, on the best-fit value of $\Omega$, avoiding volume effects.}\label{fig:laplace_example}
    \end{center}
  \end{figure*}

  \subsection{Profiling and analytical marginalisation}\label{ssec:like.approx}
    In order to approximate the marginal distribution in Eq.~\eqref{eq:marg}, let us start by considering the best-fit value of the nuisance parameters having fixed $\goodpar$. That is, we define $\badpar_*(\goodpar)$ as
    \begin{equation}
      \badpar_*(\goodpar)\equiv {\rm arg\,\,max}_{\badpar} p(\goodpar,\badpar|{\bf d}).
    \end{equation}
    Assuming that the distribution is differentiable at all points, $\badpar_*$ then satisfies
    \begin{equation}\label{eq:minim}
      \left.\frac{\partial \chi^2}{\partial\badpar}\right|_{\badpar_*}=0.
    \end{equation}

    Following \cite{2010MNRAS.408..865T}, we can then approximate the distribution at each value of $\goodpar$ by expanding $\chi^2$ to second order in $\badpar$ around $\badpar_*$, i.e.:
    \begin{equation}
      \chi^2(\goodpar,\badpar)\simeq\chi^2_*(\goodpar)+\Delta\badpar^T\mathcal{F}_*\Delta\badpar,
    \end{equation}
    where $\chi^2_*(\goodpar)\equiv\chi^2(\goodpar,\badpar_*)$, $\Delta\badpar\equiv\badpar-\badpar_*$, and we have defined the matrix
    \begin{equation}
      \mathcal{F}_{*,ij}=\left.\frac{1}{2}\frac{\partial^2\chi^2}{\partial n_i\partial n_j}\right|_{\badpar_*}.
    \end{equation}
    This is the so-called \emph{Laplace approximation} \citep{Kass1990TheVO}. In this limit, the distribution is locally (i.e. at each $\goodpar$) a multivariate normal distribution in $\badpar$, and thus the integral in Eq.~\eqref{eq:marg} can be solved analytically. The resulting marginalised likelihood has a $\chi^2_m(\goodpar)\equiv-2\log p(\goodpar|{\bf d})$ given by
    \begin{equation}\label{eq:laplace}
      \chi^2_m(\goodpar)\simeq\chi_*^2(\goodpar)+\log\left\{\det\left[\mathcal{F}_*(\goodpar)\right]\right\}+{\rm const.}.
    \end{equation}

    In what follows, we will label the two contributions in Eq.~\eqref{eq:laplace},  $\chi^2_*$ and $\log\det\mathcal{F}_*$, as the \emph{profile} and \emph{Laplace} terms respectively:
    \begin{enumerate}
      \item The {\sl profile} term is related to the ``profile likelihood'' \citep{profile1}, defined as
      \begin{equation}
        p_{\rm prof}(\goodpar|{\bf d})\propto p(\goodpar,\badpar_*|{\bf d}).
      \end{equation}
      The profile likelihood is a tool commonly used in frequentist parameter inference \citep{Cousins:1994yw}. The advantage of the profile likelihood is that its maximum is, by definition, the global maximum of the joint distribution. Understanding this maximum as an estimator for $\goodpar$ given the data, constraints on $\goodpar$ can be obtained by calculating this maximum for random simulated realisations of the data. Alternatively, if the distribution is sufficiently close to a Gaussian, these constraints can be simply obtained in terms of thresholds of the associated residual $\chi^2$, $\Delta\chi^2(\vec{\Omega})=\chi^2(\Vec{\Omega})-\chi^2_{\rm min}$ \citep{1998PhRvD..57.3873F}. Additionally, the posterior profile likelihood is, by construction, centered on the best-fit parameters, and is thus free from volume effects associated with the choice of the nuisance parameters \citep{2007JCAP...08..021H,2022ApJ...929L..16H,2022ApJ...941..110C} (see Section \ref{ssec:like.jeffrey}).
      \item The {\sl Laplace} term, sometimes referred to as {\sl Occam's razor} term, is associated with the quadratic contribution to the Laplace approximation (Eq.~\eqref{eq:laplace}), and accounts, to first order, for the volume in the space of nuisance parameters $\badpar$ that has been integrated over for fixed $\goodpar$ (i.e. the local curvature of the joint distribution at each $\goodpar$)\footnote{In the frequentist context, \cite{10.2307/2335549} introduced the formula in Eq.~\eqref{eq:laplace} under the name of ``modified profile likelihood''.}. As we will see in Section \ref{ssec:like.jeffrey}, the Laplace term is associated with volume effects, and is subdominant with respect to the profile term for sufficiently constraining data.
    \end{enumerate}
    The role of the {\sl profile} and {\sl Laplace} terms is illustrated in Fig.~\ref{fig:laplace_example}. The figure shows a bivariate distribution for two parameters with an approximate degeneracy of the form $n\Omega^{1.2}\sim{\rm const.}$. The contour levels of the true distribution are shown in red in the bottom left panel, while the black solid line shows the best-fit value of $n$ as a function of $\Omega$. The middle panel shows the exact $\chi^2$ of the distribution as a function of $n$ for a fixed $\Omega=\Omega_0$ (for convenience, we chose $\Omega_0$ to be the maximum of the distribution, also shown as a dotted line in the bottom panel). The black dashed line shows the quadratic Laplace approximation to the red curve, with the position of the best-fit $n$ (for $\Omega=\Omega_0$) marked by the blue line. The top panel shows the distribution along the $\Omega=\Omega_0$ line. The exact distribution is again shown in red, and the Laplace approximation to it is shown in dashed black. The ``profile likelihood'' approximation, which fixes $n$ to its best-fit value, is shown in blue. Finally, the bottom right panel shows the distribution marginalised over $n$, $p(\Omega)$. The true marginal is shown in red. The result of analytically marginalising over $n$ using the Laplace approximation is shown in dashed black, and recovers the true marginal almost exactly. Finally, the conditionally maximised distribution accounting only for the {\sl profile} term in Eq.~\eqref{eq:laplace} is shown in blue. As mentioned above, the profile-only approximation recovers a distribution that is centered at the best-fit value of $\Omega$ (marked by the dotted line). The shift in the peak of the true marginal observed is caused by volume effects, which we discuss in more detail in Section \ref{ssec:like.jeffrey}.

    Two qualitative results should be borne in mind in what follows. First, the Laplace approximation provides a reasonably accurate prediction for the marginal for sufficiently well-behaved distributions. Secondly, keeping only the {\sl profile} term, $p(\goodpar,\badpar_*|{\bf d})$, recovers a distribution that has approximately the same width but is, by construction, centered on the maximum of the full (un-marginalised) distribution, $p(\goodpar,\badpar|{\bf d})$.

    It is worth noting that including the Laplace term in Eq.~\eqref{eq:laplace} should come at virtually no additional computational cost. Finding $\badpar_*(\goodpar)$ requires solving for $\partial_{\badpar}\chi^2=0$, which can be done efficiently using gradient descent methods. Finding the optimal step size in these algorithms often requires evaluating the Hessian of the function being minimised, and therefore the matrix $\mathcal{F}_*$ entering the Laplace term, is already a product of the minimisation algorithm. For instance, the iteration in the case of the Newton-Raphson algorithm is given by
    \begin{equation}
      \badpar_{*,i+1}=\badpar_{*,i}-\left([\nabla_n\nabla_n^T\chi^2]^{-1}\cdot\nabla_n\chi^2\right)_i,
    \end{equation}
    where $\nabla_n\chi^2$ is the gradient of $\chi^2$ with respect to $\badpar$, and $\nabla_n\nabla_n^T\chi^2\equiv 2\mathcal{F}$ is its Hessian matrix. In the applications we will explore here, when Eq.~\eqref{eq:minim} cannot be solved analytically, we will make use of a modified version of the Newton-Raphson algorithm, which we describe in the next section.

  \subsection{Gaussian likelihoods}\label{ssec:like.gaus}
    Let us now apply the method described in the previous section to the case of Gaussian likelihoods. In this case we assume that the posterior distribution takes the form:
    \begin{equation}\label{eq:gauslike}
      -2\log p(\goodpar,\badpar|{\bf d})=({\bf d}-{\bf t})^T{\sf C}^{-1}({\bf d}-{\bf t})+\chi^2_{p,\Omega}(\goodpar)+\chi^2_{p,n}(\badpar).
    \end{equation}
    Here ${\bf t}(\goodpar,\badpar)$ is the theory vector, which depends on the model parameters, ${\sf C}$ is the covariance matrix of the data, which we assume to be model-independent, and $\chi^2_{p,\Omega}$ and $\chi^2_{p,n}$ are the parameter priors. Although the methodology described below is straightforward to generalise to the case of arbitrary priors, for simplicity we will assume that the nuisance parameters have Gaussian priors, and therefore
    \begin{equation}
      \chi^2_{p,n}(\badpar)=(\badpar-\badpar_p)^T{\sf C}_n^{-1}(\badpar-\badpar_p),
    \end{equation}
    where ${\sf C}_n$ is the prior covariance. In the case of non-Gaussian priors, it is often possible to apply a transformation to the nuisance parameters that Gaussianizes (e.g. via normalizing flows) without introducing any pathologies (singularities, etc.).

    In order to find $\badpar_*$ and $\mathcal{F}$, we need the first and second derivatives of the $\chi^2$ with respect to $\badpar$. In this case, these are given by:
    \begin{align}
      &\frac{\partial\chi^2}{\partial n_i}=-2\partial_i{\bf t}^T{\sf C}^{-1}({\bf d}-{\bf t})+2\sum_j\left[{\sf C}^{-1}_n\right]_{ij}(n_j-n_{p,j}),\\
      &\mathcal{F}_{ij}=F_{ij}+\Delta\mathcal{F}_{ij},
    \end{align}
    where we have used the shorthand $\partial_i\equiv\partial/\partial n_i$, and we have defined
    \begin{align}\label{eq:fisher}
      &F_{ij}\equiv\partial_i{\bf t}^T{\sf C}^{-1}\,\partial_j{\bf t}+\left[{\sf C}^{-1}_n\right]_{ij},\\
      &\Delta\mathcal{F}_{ij}\equiv\partial_i\partial_j{\bf t}^T\,{\sf C}^{-1}({\bf t}-{\bf d}).
    \end{align}
    
    On the one hand, the first contribution to $\mathcal{F}$, $F$, has three interesting properties:
    \begin{itemize}
        \item It is positive-definite, and therefore invertible.
        \item It is independent of the data ${\bf d}$.
        \item It coincides with the Fisher matrix of the Gaussian likelihood of Eq.~\eqref{eq:gauslike} with respect to the nuisance parameters.
    \end{itemize}
    On the other hand, when evaluated on the hypersurface $\badpar=\badpar_*(\goodpar)$, ${\bf t}$ is close to ${\bf d}$, and therefore the contribution from $\Delta\mathcal{F}$ is usually smaller than $F$. These properties will be important when discussing volume effects in the next section. For now, we will use the positive-definiteness of $F$ to define a modified Newton-Raphson iteration, by swapping the Hessian matrix $2\mathcal{F}$ with $2F$:
    \begin{equation}
      \badpar_{*,i+1}=\badpar_{*,i}-\left(\frac{1}{2}F^{-1}\,\nabla_n\chi^2\right)_i.
    \end{equation}
    This results in the Gauss-Newton method, which improves on the Newton-Raphson iteration in two aspects: first, $F$ is invertible and likely more stable than the Hessian matrix, which need not be positive-definite. Secondly, $F$ does not require calculating second derivatives of the theory vector (although this is not a computationally challenging problem for the cases considered here). Note that replacing Hessian matrix with Fisher matrix in Newton-Raphson iteration often appears in various problem, most notably in optimal quadratic estimators \citep{1997PhRvD..55.5895T}, see \cite{2015JCAP...01..022M} for further discussion. Other approaches, such as the Levenberg-Marquardt algorithm \citep{Levenberg,Marquardt}, build on this method by improving the numerical stability of $F^{-1}$ through regularisation. The problems addressed here do not require us to resort to these.

  \subsection{Volume effects and priors}\label{ssec:like.jeffrey}
    Consider now the case of a data vector with a Gaussian likelihood and poor constraining power (e.g. noise-dominated). In this limit, we would na\"ively expect the marginalised posterior distribution to return largely unconstrained $\goodpar$. To verify if this is the case, consider first the profile contribution in Eq.~\eqref{eq:laplace}, averaged over realisations of the data:
    \begin{equation}
      \langle \chi^2_*\rangle(\goodpar)= {\rm Tr}\left(\left\langle({\bf d}-{\bf t}(\goodpar,\badpar_*))({\bf d}-{\bf t}(\goodpar,\badpar_*))^T\right\rangle{\sf C}^{-1}\right).
    \end{equation}
    If the data are noise-dominated, the fluctuations in ${\bf d}-{\bf t}(\goodpar,\badpar_*)$ are dominated by noise, rather than the changes in $\goodpar$. In that case $\langle ({\bf d}-{\bf t})^T({\bf d}-{\bf t})\rangle\simeq{\sf C}$, and therefore
    \begin{equation}
      \langle\chi_*^2\rangle(\Omega)\simeq N_d,
    \end{equation}
    where $N_d$ is the number of data points. Thus, as we expected, this contribution tends to a constant that does not favour any region of parameter space.
    
    Consider now the Laplace contribution. As we argued, the contribution $\Delta\mathcal{F}_*$ is normally small compared to $F_*$ (both evaluated at $\badpar_*$), and therefore
    \begin{equation}
      \log\det\mathcal{F}_*=\log\det(F_*+\Delta\mathcal{F}_*)\simeq\log\det F_*
    \end{equation}
    which, as we discussed before, is independent of the data. The Laplace contribution to the marginal distribution is thus a parameter-dependent function that does not depend on the data, and which would favour particular regions of parameter space even in the absence of data!

    This is an example of a ``volume effect'': the process of marginalisation favours regions of parameter space that cover a larger volume of the probability density in the direction of integration, causing a shift in the maximum of the marginalised distribution with respect to the maximum of the full distribution. Volume effects depend on the definition of the nuisance parameters and, if informative, the associated priors. A classical example is trying to fit noisy data to a power-law model of the form $n\,x^\Omega$, where $x$ is an independent variable taking values $x>1$, and $(n,\Omega)$ are free parameters. If the data are noise-dominated, scattering around zero, a very negative value of the power law index $\Omega$ is able to obtain a reasonable fit to the data for a wider range of amplitudes $n$, and thus the marginalised distribution will favour values of $\Omega$ that are significantly different from the best-fit. This often happens for non-linear parameters like $\Omega$ when marginalising over amplitude-like parameters such as $n$. The most pernicious aspect of volume effects is that their impact on the marginalised distribution depends on the specific parametrisation used to define the nuisance parameters. In the example above, redefining the model to $n\,(x/x_0)^\Omega$, where $x_0$ is a fixed quantity, larger than any value of $x$, would result in marginalised posteriors that favour large and positive values of $\Omega$ for noisy data (instead of negative).

    A common choice to eliminate the dependence on model parametrisation, and thus to partially mitigate the impact of volume effects is to use the well-known \emph{Jeffreys prior}:
    \begin{equation}
      p_{\rm J}(\vec{\theta})=\sqrt{\det{F}},
    \end{equation}
    where $F$ is the Fisher information matrix, which we introduced in the previous section (Eq.~\eqref{eq:fisher} for Gaussian distributions). Comparing this with Eq.~\eqref{eq:laplace}, we can see that the inclusion of a Jeffreys prior for the nuisance parameters has the effect of partially cancelling the contribution from the Laplace term (since, as we argued, $\Delta\mathcal{F}$ is generally smaller than $F$\footnote{In addition to this, whereas $F$ is evaluated along $\badpar_*$ in the Laplace approximation, the Jeffreys prior is evaluated at every point in parameter space.}). This may not be entirely surprising since, as we saw, the Laplace term is in essence the type of volume effect that the Jeffreys prior is meant to address. However, this detour leads us to an interesting result: for Gaussian data with negligible parameter dependence in the covariance matrix, the profile likelihood will in general be a reasonable approximation to the true marginalised posterior in the presence of a Jeffreys prior. In this sense, maximisation and marginalisation are approximately equivalent as long as we avoid volume effects. We note that the strictly correct way of applying Jeffreys prior would be to compute the full Fisher matrix (i.e., for both cosmological and nuisance parameters), as our approach ignores the interdependence between the $\goodpar$ parameters, but we argue that this has but a small effect.
    
  \subsection{Linear parameters}\label{ssec:like.linear}
    Let us now consider the case of linear parameters. I.e. consider a Gaussian likelihood in the form of Eq.~\eqref{eq:gauslike} where all parameters live in the theory prediction, which has the form
    \begin{equation}\label{eq:t_linear}
      {\bf t}={\bf t}_0+{\sf T}\badpar,
    \end{equation}
    where ${\bf t}_0$ and ${\sf T}$ are a vector and a matrix independent of $\badpar$, but potentially dependent on $\goodpar$. For simplicity, we will assume that the prior on $\badpar$ is centered at zero ($\badpar_p=0$). This can always be achieved by simply redefining $\badpar'\equiv\badpar-\badpar_p$, and adding the contribution ${\sf T}\badpar_p$ to ${\bf t}_0$.

    The case of linear parameters is particularly interesting, because the $\chi^2$ is quadratic in $\badpar$ by construction, and the Laplace approximation is exact. Since the second derivatives of the theory vector are zero, $\Delta\mathcal{F}=0$, and the Fisher matrix is independent of $\badpar$ and given by
    \begin{equation}\label{eq:fisher_linear}
      \mathcal{F}=F={\sf T}^T{\sf C}^{-1}{\sf T}+{\sf C}_n^{-1}.
    \end{equation}
    Furthermore, the best-fit parameters can be found analytically:
    \begin{equation}\label{eq:lstsqr}
      \badpar_* = F^{-1}\,{\sf T}^T{\sf C}^{-1}{\bf r},
    \end{equation}
    where ${\bf r}\equiv{\bf d}-{\bf t}_0$ is the data rescaled by the $\badpar$-independent component of the theory. Using $\badpar_*$ to compute the $\chi^2$, and using Eq.~\eqref{eq:fisher_linear}, we obtain the marginalised $\chi^2_m$ of Eq.~\eqref{eq:laplace} which, as we said, is exact in this case.

    The first thing worth noting is that, if the matrix ${\sf T}$ is independent of $\goodpar$, then $F$ is constant, and so is the Laplace term. Up to an irrelevant overall constant, the marginalised $\chi^2$ is then equivalent to $\chi^2_*$, obtained by substituting the best-fit value of $\badpar$. Thus, the approximate relation between marginalisation and maximisation we outlined in the previous section becomes an equivalence when the data is Gaussian with a linear model in $\badpar$ since, in this case, there are no volume effects. All volume effects resulting from a dependence of ${\sf T}$ on $\goodpar$ are otherwise incorporated exactly in the Laplace term. If the priors on $\badpar$ are sufficiently tight, the second term in Eq.~\eqref{eq:fisher_linear} dominates, and these volume effects become negligible.

    Let us now focus on the profile term. Substituting $\badpar_*$ from Eq.~\eqref{eq:lstsqr}, we obtain
    \begin{equation}\label{eq:chi2_linear}
      \chi^2_*=({\sf W}{\bf r})^T{\sf C}^{-1}({\sf W}{\bf r})+{\bf r}^T{\sf C}^{-1}{\sf T}F^{-1}{\sf C}_n^{-1}F^{-1}{\sf T}^T{\sf C}^{-1}{\bf r},
    \end{equation}
    where the second term comes from the prior on $\badpar$, and we have defined the matrix
    \begin{equation}
      {\sf W}\equiv{\sf I}-{\sf T}F^{-1}{\sf T}^T{\sf C}^{-1},
    \end{equation}
    with ${\sf I}$ the identity.
    
    First, consider the limit of no external prior (i.e. ${\sf C}_n^{-1}=0$). In this case, we can ignore the second term in Eq.~\eqref{eq:chi2_linear}, and the Fisher matrix is $F={\sf T}^T{\sf C}^{-1}{\sf T}$. We can then see that the matrix ${\sf W}$ projects ${\bf r}$ onto the subspace that is orthogonal to all the columns of ${\sf T}$ (with orthogonality defined using the inverse covariance of the data ${\sf C}^{-1}$ as a dot product). Marginalising over linear parameters is therefore equivalent in this limit to deprojecting all modes of the data that live in the subspace spanned by the columns of ${\sf T}$ \citep{1992ApJ...398..169R}.

    Secondly, Eq.~\eqref{eq:chi2_linear} can be simplified significantly into
    \begin{equation}
      \chi^2_*={\bf r}^T\tilde{\sf C}^{-1}{\bf r},
      \label{eq:cov}
    \end{equation}
    where $\tilde{\sf C}$ is a modified covariance given by
    \begin{equation}\label{eq:modcov}
      \tilde{\sf C}={\sf C}+{\sf T}{\sf C}_n{\sf T}^T.
    \end{equation}
    To obtain this beautifully simple result, one only needs to expand first term in Eq.~\eqref{eq:chi2_linear}, simplify the result, and make use of the Woodbury matrix identity \citep{woodbury}. It is worth stressing again that this result is an exact expression for both the marginal posterior (using a Jeffreys prior) and the conditionally maximised posterior, as explained in the previous section. Maximising and marginalising over $\badpar$ therefore result in the same Gaussian likelihood with the theory vector evaluated at $\badpar=0$ (or at its prior mean if non-zero), and a modified covariance $\tilde{\sf C}$, obtained by simply assigning additional variance in quadrature to the modes of the data that align with the columns of ${\sf T}$ (with this extra variance given by the associated $\badpar$ parameter priors).

    Note that while this approach is algorithmically the fastest, it does not produce a best-fit value of a given nuisance parameter. This is occasionally useful even for nuisance parameters. A canonical example is the shot noise contribution to the galaxy power spectrum, the value of which can tell us about the halos in which the given tracer galaxies reside.

    To summarise: in the case of Gaussian data, negligible parameter dependence of the covariance matrix, and a theory model that is linear in the nuisance parameters, the Laplace approximation is exact. In this case, there is a mathematical equivalence between marginalisation (using a Jeffreys prior), $\chi^2$ minimisation, deprojection, and simply adding in quadrature the prior uncertainty on the marginalised parameters at the data level. If the modes associated with $\badpar$ (i.e. the columns of ${\sf T}$) depend on the other parameters of the model, the associated volume effects are captured exactly by the Laplace term, which is simply given by the log-determinant of Eq.~\eqref{eq:fisher_linear}. Importantly, in this scenario, volume effects become negligible if the constraints on nuisance parameters are either dominated by data, or by the priors.

\section{Cosmology from tomographic large-scale structure data}\label{sec:tomo}
  To explore the validity of the Laplace approximation introduced in the previous section to marginalise over nuisance parameters in the context of cosmology, we will study the case of 3$\times$2pt analyses, combining photometric galaxy clustering and weak lensing.

  \subsection{Galaxy clustering and weak lensing}\label{ssec:tomo.3x2pt}
    Photometric redshift surveys make use of two main cosmological probes: cosmic shear (i.e. the distortion in the shape of galaxies caused by weak gravitational lensing) and galaxy clustering. In both cases, the data are typically split into photometric redshift bins, and the data vector is constructed by combining various angular auto- and cross-correlations between pairs of such bins. The galaxy overdensity $\delta_g^\alpha(\nv)$ and the weak lensing shear $\gamma^\alpha_G(\nv)$ for galaxies in redshift bin $\alpha$ can be related to the three-dimensional fluctuations in the galaxy number density $\Delta_g({\bf x})$ and the matter density $\Delta_m({\bf x})$ via \citep{2001PhR...340..291B,2017arXiv170609359K}
    \begin{align}\nonumber
      &\delta_g^\alpha(\nv)=\int_0^{\chi_H}d\chi\,q^\alpha_g(\chi)\,\Delta_g(\chi(z)\nv, z),\hspace{12pt}\\
      &\gamma^\alpha_G(\nv)=\int _0^{\chi_H}d\chi\,q^\alpha_G(\chi)\,\left[-\chi^{-2}\eth^2\nabla^{-2}\Delta_m(\chi\nv,z)\right],
      \label{eq:deltagamma} 
    \end{align}
    where $\nv$ is the sky direction, $\chi$ is the comoving radial distance at redshift $z$, $\chi_H$ is the distance to the horizon, $q^\alpha_g$ and $q^\alpha_\gamma$ are the radial kernels for galaxy clustering and cosmic shear, respectively, and $\eth$ is the spin-raising differential operator, acting on a spin-$s$ quantity as:
    \begin{equation}
      \eth\,_sf(\theta,\varphi)=-(\sin\theta)^s\left(\frac{\partial}{\partial\theta}+\frac{i}{\sin\theta}\frac{\partial}{\partial\varphi}\right)(\sin\theta)^{-s}\,_sf
    \end{equation}
    and turning it into a spin-$(s+1)$ quantity.
    
    The radial kernels in both cases are given by
    \begin{align}\nonumber
      &q^\alpha_g(\chi)\equiv\frac{H(z)}{c}p_\alpha(z),\\
      &q^\alpha_G(\chi)\equiv\frac{3}{2}H_0^2\Omega_m\frac{\chi}{a(\chi)}\int_{z(\chi)}^\infty dz' p_\alpha(z')\frac{\chi(z')-\chi}{\chi(z')},
    \end{align}
    where $c$ is the speed of light, $H(z)$ is the Hubble expansion rate, $H_0\equiv H(z=0)$, $\Omega_m$ is the matter density parameter today, and $p_\alpha(z)$ is the redshift distribution in bin $\alpha$,

    Cosmic shear observations are sensitive not only to the weak lensing distortion of galaxy shapes, but also to the intrinsic alignments (IAs) in the orientation of galaxies due to local interactions. The total observed cosmic shear signal is thus
    \begin{equation}\label{eq:gamma}
      \gamma^\alpha=\gamma^\alpha_G+A_{\rm IA}\gamma^\alpha_I,
    \end{equation}
    where $\gamma^\alpha_G$ is the lensing signal, given by Eq.~\eqref{eq:deltagamma}, and $A_{\rm IA}\gamma^\alpha_I$ is the intrinsic alignment component. Using the linear non-linear alignment model, the latter is given by
    \begin{equation}
      \gamma^\alpha_I(\nv)=-\int_0^{\chi_{H}}d\chi\,q_I^\alpha(\chi)[-\chi^{-2}\eth^2\nabla^{-2}\Delta_m(\chi\nv,z)],
    \end{equation}
    and $A_{\rm IA}$ is an unknown amplitude parameter describing the strength of the local alignment signal. The IA radial kernel is the same as the galaxy clustering kernel
    \begin{equation}
      q_I^\alpha(\chi)=\frac{H(z)}{c}p_\alpha(z).
    \end{equation}

    The relation between the galaxy overdensity $\Delta_g$ and the matter overdensity $\Delta_m$ is complex in detail (non-linear, non-local and stochastic). Over mildly non-linear scales, one can however use a perturbative approach, relating $\Delta_g$ with scalar combinations of the Hessian of the gravitational potential $\partial_i\partial_j\Phi$ (where $\Phi$ is normalised so that $\nabla^2\Phi\equiv\Delta_m$). Following \citet{2009JCAP...08..020M,2018JCAP...07..029A}, here we will expand this bias relation to second order, including non-local contributions, so that:
    \begin{equation}\label{eq:bias_pt} 
      \Delta_g=b_1\Delta_m+\frac{b_2}{2!}(\Delta_m^2-\langle\Delta_m^2\rangle)+\frac{b_s}{2!}(s^2-\langle s^2\rangle)+b_\nabla \nabla^2\Delta_m.
    \end{equation}
    Here $s^2\equiv s_{ij}s^{ij}$ is the trace of the squared tidal tensor, where $s_{ij}\equiv\partial_i\partial_j\Phi-\nabla^2\Phi/3$. The quantities $b_1$, $b_2$, $b_s$, and $b_\nabla$ are the so-called ``linear'', ``quadratic'', ``tidal'', and ``non-local'' bias parameters, which characterise the response of the galaxy overdensity to the corresponding terms in perturbation theory, including the impact of non-local effects on scales comparable with the Lagrangian halo size. Within this formalism, and assuming that the bias parameters are constant within each redshift bin, the projected galaxy overdensity can then be expressed as a sum over the projected version of the different operators in tishe previous equation:
    \begin{equation}\label{eq:deltag}
      \delta_g^\alpha(\nv)=\sum_k b_{\alpha,k},\delta_k^\alpha(\nv),
    \end{equation}
    where $k$ runs over the set $\{1,2,s,\nabla\}$, corresponding to the various operators, dependent only on the matter overdensity and tidal tensor, that contribute to the total galaxy overdensity. $b_{\alpha,k}$ is the value of each associated bias parameter in bin $\alpha$, and 
    \begin{equation}
      \delta_k^\alpha(\nv)\equiv\int_0^{\chi_H}d\chi\,q_g^\alpha(\chi)\Delta_k(\nv),
    \end{equation}
    with
    \begin{align}\nonumber
      &\Delta_1\equiv\Delta_m,\hspace{6pt}\Delta_2\equiv\Delta_m^2-\langle\Delta_m^2\rangle,\\\label{eq:bias_obs}
      &\Delta_s\equiv s^2-\langle s^2\rangle,\hspace{6pt}\Delta_\nabla\equiv\nabla^2\Delta_m.
    \end{align}
    For some of the results shown in Section \ref{sec:res} we will consider only linear bias, setting $b_2=b_s=b_\nabla=0$. Either of these bias schemes is only valid on sufficiently large scales, and therefore we will limit our analysis to multipoles $\ell<k_{\rm max}\bar{\chi}$, where $\bar{\chi}$ is the comoving distance to the mean redshift of the galaxy tracer under analysis. The maximum wavenumber used will be $k_{\rm max}=0.15\,{\rm Mpc}^{-1}$ when using linear bias, and $k_{\rm max}=0.3\,{\rm Mpc}^{-1}$ when using the perturbative approach above \citep{2020PhRvD.102l3522P}.

    Comparing Eq.~\eqref{eq:gamma} and Eq.~\eqref{eq:deltag}, we see that we can describe both tomographic galaxy clustering and cosmic shear as a projected tracer $u^\alpha(\nv)$
    with the generic form
    \begin{equation}\label{eq:generic_tracer}
      u^\alpha(\nv)=\epsilon_u\,u_M^\alpha(\nv)+\sum_k b^u_{\alpha,k}\,u_k^\alpha(\nv)\, ,
    \end{equation}
    where $b^u_{\alpha,k}$ are bias parameters (specifying the tracer type $u$), $u_M^\alpha$ and $u_k^\alpha$ are projected quantities that depend only on cosmological observables (matter overdensities, comoving distances etc.) and the radial kernels, and $\epsilon_u$ is a Boolean variable that is either 1 if the tracer contains an unbiased contribution (as is the case of cosmic shear), and 0 otherwise (as is the case of galaxy clustering). The index $\alpha$ in Eq.~\eqref{eq:generic_tracer} runs over the redshift bins, which allows for the general case of having redshift-dependent bias functions. If this is not the case, $b^u_{\alpha,k}\equiv b^u_k$ can optionally be assumed to not vary across redshift bins.

    We can relate the power spectrum, $P_{XY}(k,z)$, of two three-dimensional quantities $X$ and $Y$ (e.g., $\Delta_k$) to the angular cross-power spectrum of their associated projected tracers in redshift bins $\alpha$ and $\beta$ (e.g. $\delta_k^\alpha$) via:
    \begin{equation}\label{eq:limber}
      C_\ell^{(X,\alpha),(Y,\beta)}=\int\frac{d\chi}{\chi^2}\,q_X^\alpha(\chi)\,q_Y^\beta(\chi)\,P_{XY}\left(k=\frac{\ell+1/2}{\chi},z(\chi)\right),
    \end{equation}
    where we have assumed the Limber approximation \citep{1953ApJ...117..134L,2004PhRvD..69h3524A}, appropriate for the wide radial kernels considered in this work. Note that in principle the lensing kernel should be multiplied by an $\ell$-dependent prefactor
    \begin{equation}
      G_\ell\equiv\sqrt{\frac{(\ell+2)!}{(\ell-2)!}}\frac{1}{(\ell+1/2)^2}
    \end{equation}
    to account for the difference between angular and three-dimensional derivatives in Eq.~\eqref{eq:deltagamma} (i.e. $\chi^2\eth^2\nabla^{-2}\not\equiv1$).

    To calculate the matter power spectrum we will use the {\tt Halofit} fitting function \cite{2003MNRAS.341.1311S} with revisions from \cite{2012ApJ...761..152T}. To calculate the power spectra between the different perturbative matter fields involved in the bias expansion (Eq.~\eqref{eq:bias_obs}) we will make use of {\tt Fast-PT} \footnote{\url{https://github.com/JoeMcEwen/FAST-PT}}. The procedure is described in \citet{2016JCAP...09..015M} and we refer readers to that paper for further details.

    The cosmological analysis of tomographic weak lensing and galaxy clustering two-point functions requires accounting for, and propagating uncertainties in some of the ingredients of the corresponding theoretical predictions. This is usually done by defining a model that describes the impact of the associated effects, and marginalising over the nuisance parameters of that model. The nature of these parameters can be classified into two main types:
    \begin{itemize}
      \item Calibratable parameters: these are parameters on which relatively tight priors can be placed using external data (i.e. they can be calibrated). These are normally associated with observational effects, such as shape measurement or photometric redshift errors.
      \item Non-calibratable parameters: these are parameters for which no reliable prior information exists, and which must therefore be measured with our own data. These are normally associated with astrophysical uncertainties specific to the sample under study, such as galaxy bias or intrinsic alignment parameters.
    \end{itemize}
    The next two sections describe the strategies we will use to marginalise over both types of nuisance parameters.

  \subsection{Calibratable systematics: linearisation}\label{ssec:tomo.linear}
    In the presence of tight priors, which we will further assume to be Gaussian, the nuisance parameters may not stray far from their prior mean. In that case, we can Taylor-expand the theory prediction as in Eq.~\eqref{eq:t_linear}
    \begin{equation}
      {\bf t}(\goodpar,\badpar)={\bf t}_0(\goodpar)+{\sf T}\,(\badpar-\badpar_p),
    \end{equation}
    where $\badpar_p$ is the prior mean, and
    \begin{equation}
      {\bf t}_0\equiv{\bf t}(\goodpar,\badpar_p),\hspace{6pt}
      {\sf T}\equiv\left.\frac{\partial{\bf t}}{\partial\badpar}\right|_{\badpar_p}.
    \end{equation}
    
    Since now the theory is linear with respect to $\badpar-\badpar_p$, we can then follow the procedure in Section \ref{ssec:like.linear} to analytically marginalise over those parameters. As we discussed, we simply modify the covariance of the data vector (in this case, a collection of power spectra) as in Eq.~\eqref{eq:modcov}, and then sample the resulting Gaussian likelihood evaluating ${\bf t}$ at the prior mean of $\badpar$. Two important things should be noted. First, in doing this, we have neglected the parameter dependence on $\goodpar$ of the Laplace term, given by
    \begin{equation}
      \log\left[\det\left({\sf T}^T{\sf C}^{-1}{\sf T}+{\sf C}_n^{-1}\right)\right]
    \end{equation}
    and therefore omit it from the calculation, as it only adds a multiplicative constant in this approximation. If the prior is sufficiently tight, this term is dominated by the constant ${\sf C}_n^{-1}$ contribution, so the approximation is reasonable. Secondly, since the modified covariance matrix (Eq.~\eqref{eq:modcov}) now involves a term of the form ${\sf T}{\sf C}_n{\sf T}^T$, in principle it depends on $\goodpar$ through ${\sf T}$. Calculating the covariance at each point in the likelihood may be computationally costly, depending on the size of ${\sf T}$. Instead, we will simply evaluate ${\sf T}$ at the best-fit value of $\goodpar$, and ignore all parameter dependence on $\goodpar$ of the covariance. It was shown in \citet{2019OJAp....2E...3K} that the parameter dependence of the covariance can generally be neglected and, furthermore, for a sufficiently tight prior the ${\sf T}{\sf C}_n{\sf T}^T$ contribution should be subdominant. \citet{2020JCAP...10..056H} also showed that the choice of fiducial $\goodpar$ does not affect the final results as long as they are close to the center of the posterior distribution. Adopting these two approximations (in addition to the Taylor expansion of ${\bf t}$) is therefore well justified and, as we will show in Section \ref{ssec:res.dz}, leads to accurate results.

    One of the most important calibratable systematics in photometric surveys comes from the uncertainties in the redshift distribution of the source and lens samples. To account for these, it is common to parametrise the potential deviations from a fiducial redshift distribution, and to calibrate those parameters through external data and simulations. In a simplified model, errors in photo-$z$'s cause shifts in the means and changes in the width of the derived redshift distribution for a population of galaxies \citep{2016PhRvD..94d2005B}. The resulting model for the redshift distribution is
    \begin{equation}
      p_\alpha(z) \propto \hat{p}_\alpha(z^\alpha_c + w^\alpha_{z}(z-z^\alpha_{c}) + \Delta z^\alpha),
      \label{eq:photo-z-model}
    \end{equation}
    where $\Delta z^\alpha$ and $w^\alpha_z$ parametrise deviations in the mean and width of the fiducial distribution $\hat{p}_\alpha$, and $z_c^\alpha$ is the redshift at which this fiducial distribution attains its maximum. We will label these the ``shift'' and ``width'' parameters respectively in what follows. Note that the normalisation of the distribution changes with both of these parameters, and it must be renormalised to unit area before using it to compute any power spectra.

    In the rest of this paper, we will add one shift and one width parameter for each galaxy clustering bin, and one shift parameter for each cosmic shear bin, fixing their width. On the one hand, mean shifts have a strong impact on the weak lensing kernel, since it depends on a cumulative integral of the redshift distribution, and may also affect galaxy clustering by changing the growth factor. On the other hand, mild ($\sim10\%$) changes to the width have a very strong impact on the amplitude of the clustering auto-correlations, but leave the weak lensing kernel almost unchanged \citep{RZP}.

    Including these free parameters amounts to burdening the MCMC sampler with tens of extra parameters, which inevitably leads to a substantial slowdown of its convergence. Moreover, there is no guarantee that all the uncertainty in the $p(z)$s can be captured by these parameters, and several alternative procedures have been developed to ensure this. Although we will focus here on the shift-width parametrisation, \citet{2020JCAP...10..056H} and \cite{RZP} explore the most general case of treating the $p(z)$ as a step-wise function, with the function values at each step treated as free parameters, and show that the approximate marginalisation described here leads to accurate results for both clustering and shear. We will therefore focus here only on the simpler shift-width parametrisation, to exemplify the performance of the Laplace approximation in the case of calibratable nuisance parameters.

    We stress that we can apply the same procedure to any other parameter that appears to behave linearly in the theoretical model. This is the case for any other nuisance parameter with a sufficiently tight prior, and in fact this procedure is routinely used for multiplicative bias parameters in cosmic shear analyses \citep{2020A&A...633A..69H}. We note that the general procedure of using the Laplace approximation is exact in the case of truly linear parameters, regardless of their priors. A good example of this is the amplitude of shot noise or stochastic contributions to galaxy clustering auto-correlations \citep{2105.12108}.

  \subsection{Non-calibratable systematics: bias parameters}\label{ssec:tomo.bias}
    Most sources of astrophysical uncertainty cannot be well-constrained from external data, and thus must be constrained at the same time as the cosmological parameters, and marginalised over. In this case, the linearisation described in the previous section is not appropriate, and we must resort to numerical methods in order to obtain $\badpar_*$ and the Laplace contribution in Eq.~\eqref{eq:laplace}. In the model introduced in Section \ref{ssec:tomo.3x2pt}, these astrophysical uncertainties are described by the bias and intrinsic alignment parameters. In this formalism, all tracers can be expressed generically as in Eq.~\eqref{eq:generic_tracer}, where $u_M^\alpha$ and $u_k^\alpha$ are projected maps depending purely cosmological fields (i.e. depending only on the matter overdensity and the tidal field), and all astrophysical uncertainties are incorporated in the $b^u_{\alpha,k}$ parameters.

    Although the bias/IA description used here covers a wide range of state-of-the-art physical models used in current 3$\times$2pt analyses, it is mathematically exceptionally simple. From Eq.~\eqref{eq:generic_tracer} we see that the cross-correlation between any two such tracers ($u^\alpha$, $w^\beta$) is a simple quadratic function of the bias parameters:
    \begin{align}\nonumber
        C_\ell^{u^\alpha,w^\beta} &= 
        \epsilon_u \epsilon_w C_\ell^{u_M^\alpha,w_M^\beta}
        + \sum_i b^u_{\alpha,i} \epsilon_w C_\ell^{u_i^\alpha,w_M^\beta} \\\label{eq:cl_bias_model}
        &+ \sum_j \epsilon_u b^w_{\beta,j} C_\ell^{u_M^\alpha,w_j^\beta}
        + \sum_{i,j} b^u_{\alpha,i} b^w_{\beta,j} C_\ell^{u_i^\alpha, w_j^\beta}\, .
    \end{align}
    Here, $C_\ell^{u_{M/i}^\alpha,w_{M/j}^\beta}$ are the power spectra between the cosmological projected fields $u^\alpha_{M/i}$ and $w^\beta_{M/j}$, defined in Eq.~\eqref{eq:generic_tracer}, and the sums run over the associated bias terms as introduced in Eq.~\eqref{eq:deltag}. Since these only involve radial projections of purely cosmological quantities, they can be treated as templates that only depend on the cosmological parameters. Note that, in principle, these templates also depend on the calibratable nuisance parameters described in the previous section (e.g. through the modification in the radial kernels due to $p(z)$ uncertainties). However, we assume that we have been able to marginalise over these analytically as we described above, and therefore they can be treated as fixed for all intents and purposes.
    
    The first derivative of the power spectrum with respect to the bias parameters is thus a linear polynomial:
    \begin{align}\nonumber
      \frac{\partial C_\ell^{u^\alpha,w^\beta}}{\partial b_{\gamma,k}^r} =
      & \,\,\delta^K_{\alpha,\gamma}\delta^K_{u,r}\left[\epsilon_w C_\ell^{u_k^\alpha,w_M^\beta}+\sum_j b_{\beta,j}^w C_\ell^{u_k^\alpha,w_j^\beta} \right] \\
      &+ \delta^K_{\beta,\gamma} \delta^K_{w,r} \left[\epsilon_u C_\ell^{u_M^\alpha,w_k^\beta}+\sum_i b^u_{\alpha,i} C_\ell^{u_i^\alpha,w_k^\beta} \right]\, ,
    \end{align}
    where $\delta^K$ is the Kronecker delta, and $b_{\gamma,k}^r$ denotes the $k$th bias term of tracer type $r$ (i.e., galaxy overdensity or shear) in redshift bin $\gamma$. Finally, the Hessian is constant with respect to the bias parameters
    \begin{align}
      & \frac{\partial^2C^{u^\alpha,w^\beta}_\ell}{\partial b_{\gamma,k}^r\partial b_{\sigma,m}^s} =\\ \nonumber 
      & \quad  \bigg[ \delta^K_{\gamma,\alpha}\delta^K_{\sigma,\beta} \delta^K_{r,u}\delta^K_{s,w} + \delta^K_{\gamma,\beta}\delta^K_{\sigma,\alpha} \delta^K_{r,w}\delta^K_{s,u} \bigg] C_\ell^{r_k^\gamma, s_m^\sigma} \, .
    \end{align}
    
    These expressions are remarkably simple and fast to evaluate, and thus computing the $\chi^2$ and its derivatives (needed for minimisation, and to calculate the Laplace contribution) can be done extremely efficiently. As we will see, in practice we find that finding the minimum of the $\chi^2$ takes $O(10-100)$ Gauss-Newton iterations, each of which is orders of magnitude faster than recomputing the power spectrum templates when changing cosmological parameters. Computing the Laplace approximation to the marginalised posterior at each sample of the cosmological parameters is therefore virtually equivalent to evaluating the joint posterior for new cosmological $+$ nuisance parameters if using brute-force marginalisation.

  \subsection{Example Stage-III and Stage-IV datasets}\label{ssec:tomo.data}
    To test the validity of the methodology described above in practice, we will apply it to two different datasets:
    \begin{itemize}
      \item Real data from the Dark Energy Survey Year-1 (\desyo) data release \citep{2018PhRvD..98d3526A}.
      \item A simulated $3\times2$-point data vector mimicking the characteristic of a Stage-IV survey such as LSST.
    \end{itemize}
    These datasets are representative of both the current and future generation and thus span the plausible accuracy range for the foreseable future.
    We describe them next.
    
    \subsubsection{The \desyo data}\label{sssec:tomo.data.des}
      We use the galaxy-galaxy, galaxy-shear, and shear-shear power spectra and covariance matrix provided in \cite{2105.12108}, constructed from the \desyo data. DES is a five-year photometric survey which has observed 5000 deg$^2$ of the sky using five different filter bands (\textit{grizY}) from the 4m Blanco Telescope at the Cerro Tololo Inter-American Observatory (CTIO), in Chile. \cite{2105.12108} employed the publicly available key {\tt Y1KP} catalogs\footnote{\url{https://des.ncsa.illinois.edu/releases/y1a1/key-catalogs}}, covering 1786 deg$^2$ before masking \citep{2018PhRvD..98d3526A,1708.01531}.
      
      The galaxy clustering sample consists of luminous red galaxies (LRGs) selected with the \redmagic{} algorithm, divided into 5 redshift bins, defined in \cite{1708.01536}. In this analysis, we will make use of the fiducial redshift distributions released by DES to model the angular power spectra. We will also adopt the same galaxy weights to correct for sky systematics \cite[see][for more details]{1708.01536}.
    
      The shear sample is the official source sample used in the \desyo analysis \citep{1708.01533}, including all cuts and tomographic bin definitions. Galaxy shapes were determined using the \mcal{} algorithm \citep{1702.02600,1702.02601}. The sample is divided into four tomographic bins, for which we use the official redshift distributions provided with the Y1 release \citep{1708.01532}. See \citet{2010.09717} for further details regarding the estimation of the shear power spectra and covariance.
      
      Following the official \desyo analysis, we use all cross-correlations between different shear bins and between shear and clustering bins, but only the auto-correlations between clustering bins.

    \subsubsection{Synthetic Stage-IV data}\label{sssec:tomo.data.lsst}
      We consider a futuristic, idealised data set that resembles the characteristics of LSST. It is important to test our method in the low-noise regime, where the inferred posterior is even more sensitive to redshift distribution uncertainties or, in general, degeneracies between cosmological and nuisance parameters, and where the final error budget is more dominated by these effects.

      To define the clustering and shear samples we follow the same procedure outlined in \citet{bch}. The shear sample is defined following the LSST Science Requirements Document \citep{2018arXiv180901669T} (see Appendices D1 and D2). We divide this redshift distribution into 5 bins in photometric redshift space, each containing the same number of sources. We assume a Gaussian photometric redshift uncertainty with standard deviation $\sigma_z=0.05(1+z)$, which thus defines the true-redshift tails of the distribution in each tomographic bin. The sample has an overall angular number density of 27 ${\rm gals.}\,{\rm arcmin}^{-2}$. For galaxy clustering, we define a sample extending out to $z\sim1.5$ with a total density of $4\,{\rm gals.}\,{\rm arcmin}^{-2}$ (as would be expected of an LRG-like sample for LSST). This number density and the associated redshift distribution were estimated using measurements of the luminosity function for red galaxies as described in \citet{2015ApJ...814..145A}. The sample was divided into 6 redshift bins equi-spaced in photometric redshift space, and assuming a photometric redshift uncertainty of $\sigma_z=0.02(1+z)$. To simplify the analysis, we assume a constant linear galaxy bias $b_1=1$, and set all higher-order bias coefficients to zero. The results obtained in the next section should be largely insensitive to this choice.

      For simplicity, we use a Gaussian covariance to describe the uncertainties of the resulting data vector, calculated assuming a sky fraction $f_{\rm sky}=0.4$. The LSST data vector was generated assuming a true cosmology with parameters
      \begin{equation}\label{eq:cospar}
        (\Omega_m,\Omega_b,h,n_s,\sigma_8)=(0.3,\,0.05,\,0.7,\,0.96,\,0.8),
      \end{equation}
      where $\Omega_m$ and $\Omega_b$ are the total matter and baryon fractions, $h$ is the reduced Hubble parameter, $n_s$ is the scalar spectral index, and $\sigma_8$ is the standard deviation of linear density perturbations smoothed on spheres of radius $8\,{\rm Mpc}\,h^{-1}$ at redshift $z=0$.

  \subsection{Likelihood}\label{ssec:tomo.like}

\begin{table}
\centering
\def\arraystretch{1.2}
\begin{tabular}{|ll|ll|}
\hline
\multicolumn{4}{|c|}{\textbf{Parameter priors}} \\
\hline
Parameter &  Prior & Parameter &  Prior\\  
\hline 
\multicolumn{2}{|c|}{\textbf{Cosmology}}  &        \multicolumn{2}{|c|}{\textbf{Redshift calibration}} \\
$\Om$  &  $U (0.07, 0.8)$          & $\Delta z_{{\rm s}}^1 $  & $\cN (\Delta z^1_{s,*},0.016)$ \\ 
$\Ob$  &  $U (0.03, 0.07)$         & $\Delta z_{{\rm s}}^2 $  & $\cN (\Delta z^2_{s,*},0.013)$  \\
$h$   &  $U (0.55, 0.91)$          & $\Delta z_{{\rm s}}^3 $  & $\cN (\Delta z^3_{s,*}, 0.011)$ \\
$\ns$ & $U (0.87, 1.07)$           & $\Delta z_{{\rm s}}^{4,5} $  & $\cN (\Delta z^{4,5}_{s,*}, 0.022)$ \\ 
$\sig$ &  $U(0.5, 1.1)$            & $\Delta z_{\rm g}^1$ & $\cN (\Delta z^1_{g,*}, 0.007)$ \\ \cline{1-2}
\multicolumn{2}{|c|}{\textbf{Bias parameters}} & $\Delta z_{\rm g}^2$ & $\cN (\Delta z^2_{g,*}, 0.007)$ \\
$b^i_1$ & $\cN(1.5, 100)$            & $\Delta z_{\rm g}^3$ &  $\cN (\Delta z^3_{g,*}, 0.006)$ \\
$b^i_{2,s,\nabla}$ & $\cN(0, 100)$ & $\Delta z_{\rm g}^{4,5,6}$ &  $\cN (\Delta z^{4,5,6}_{g,*}, 0.01)$ \\  \cline{3-4}
$A_\mathrm{IA,0} $ & $\cN(0, 100)$ & $w^i_{z, {\rm g}}$ & $\cN (1.00, 0.08)$ \\

\hline
\end{tabular}
\caption{Prior distributions for the nuisance parameters entering our ``3$\times$2pt'' analysis for each tracer. $U(a, b)$ and $\cN(\mu, \sigma)$ describe a uniform distribution with boundaries $(a, b)$ and a Gaussian distribution with mean $\mu$ and variance $\sigma$, respectively.  The index $i$ in $b^i_\mathrm{g}$ and $m^i$ runs over the different redshift bins. $\Delta z_{*}$ denotes the deviation from zero of the central/best-fit value of each redshift uncertainty parameter.
}
\label{tab:priors_3x2pt}
\end{table}

    To obtain constraints on cosmological parameters from the real and synthetic data described in the previous section, we will assume that the data vector (i.e. the clustering and shear power spectra), follows a Gaussian likelihood as in Eq.~\eqref{eq:gauslike}, with a parameter-independent covariance. The model will be described by 5 cosmological parameters, listed in Eq.~\eqref{eq:cospar}, one or four linear parameters for each clustering redshift bins (for linear and PT bias respectively), one intrinsic alignment amplitude, one redshift shift parameter for each clustering and cosmic shear bin, and one redshift distribution width parameter for each clustering bin. The priors used for all parameters are provided in Table \ref{tab:priors_3x2pt}. The priors on cosmological parameters are roughly based on the choices made for the \desyo analysis, except we sample over $\sigma_8$ instead of the scalar spectrum amplitude $A_s$. The priors on the redshift shift parameters are based on the calibration of the \desyo data \citep{1708.01532}, and thus represent achievable calibration levels. The priors on the redshift width parameters are commensurate with those used in the DES Year-3 analysis \citep{2022PhRvD.105b3520A} (the \desyo analysis did not introduce width parameters). Finally, we place an uninformative Gaussian prior on all the bias parameters, centered at zero and with a standard deviation of 100. The choice of using a very broad Gaussian prior as opposed to simply a flat prior is intended to enforce a smooth distribution as a function of these parameters, and to potentially aid the Gauss-Newton iterator when minimising the $\chi^2$.

    We employ the \texttt{cobaya} MCMC sampler \citep{2019ascl.soft10019T,2021JCAP...05..057T} with a convergence condition that the Gelman-Rubin diagnostic, $R$, ought to satisfy $R-1 < 0.01$. When using the Laplace approximation, we minimise the $\chi^2$ over the nuisance parameters using a Gauss-Newton iterator, using the analytical derivatives with respect to bias parameters as described in Section \ref{ssec:tomo.bias}, and modify the log-probability to be sampled by \texttt{cobaya} to be that of Eq.~\eqref{eq:laplace}. We find that, in order to reduce the number of steps taken by the Gauss-Newton iterator, it is useful to determine a well educated global best-fit for the full parameter space before taking any samples, and to start the iterator from the corresponding best-fit value of the nuisance parameters.
 
    Throughout, we made use of the fitting formula of \cite{1998ApJ...496..605E} to calculate the linear matter power spectrum. We do this to speed up the calculations, and we have verified that the results obtained on the \desyo data are insensitive to this choice compared to using a Boltzmann solver such as {\tt CLASS} \citep{1104.2933}. The non-linear matter power spectrum is then computed using HALOFIT \citep{1208.2701}.

\section{Results}\label{sec:res}
  \subsection{Calibratable nuisance parameters}\label{ssec:res.dz}
    \begin{figure}
      \centering
      \includegraphics[width=0.48\textwidth]{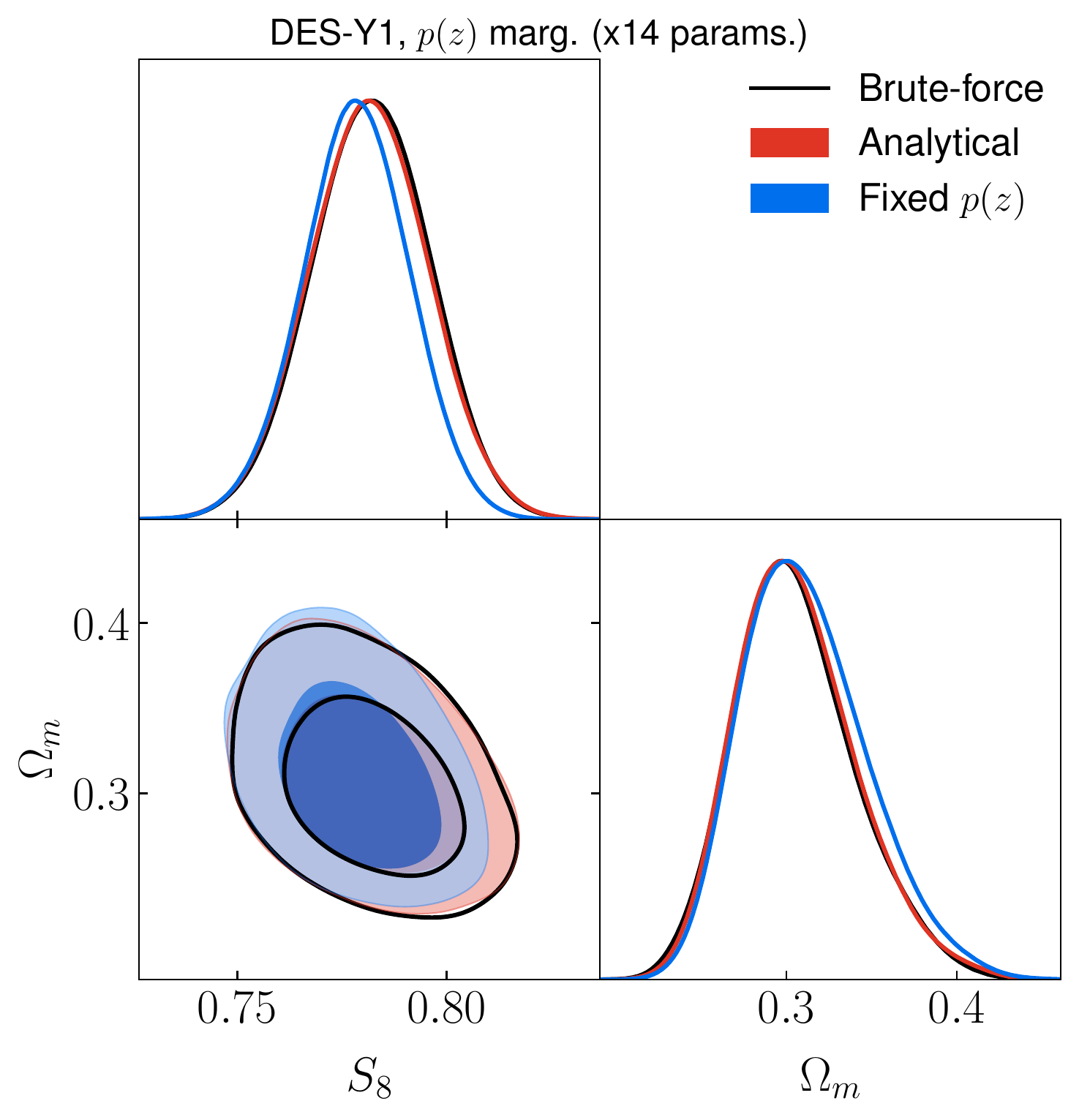}
      \caption{Contours comparing brute-force (silver) with analytic marginalisation over photo-$z$ uncertainties (red; see Section~\ref{ssec:tomo.linear}). Results are shown for the \desyo data. We find that the contours are virtually unchanged, demonstrating the benefit of using an efficient analytic marginalisation scheme. We also show for posterity the result of assuming negligible error on the photo-$z$ distributions (blue). In reality, this assumption does not hold and potentially leads to a bias in the inferred cosmology.}
      \label{fig:DES_margdzwz}
    \end{figure}
    \begin{figure*}
      \centering
      \includegraphics[width=0.6\textwidth]{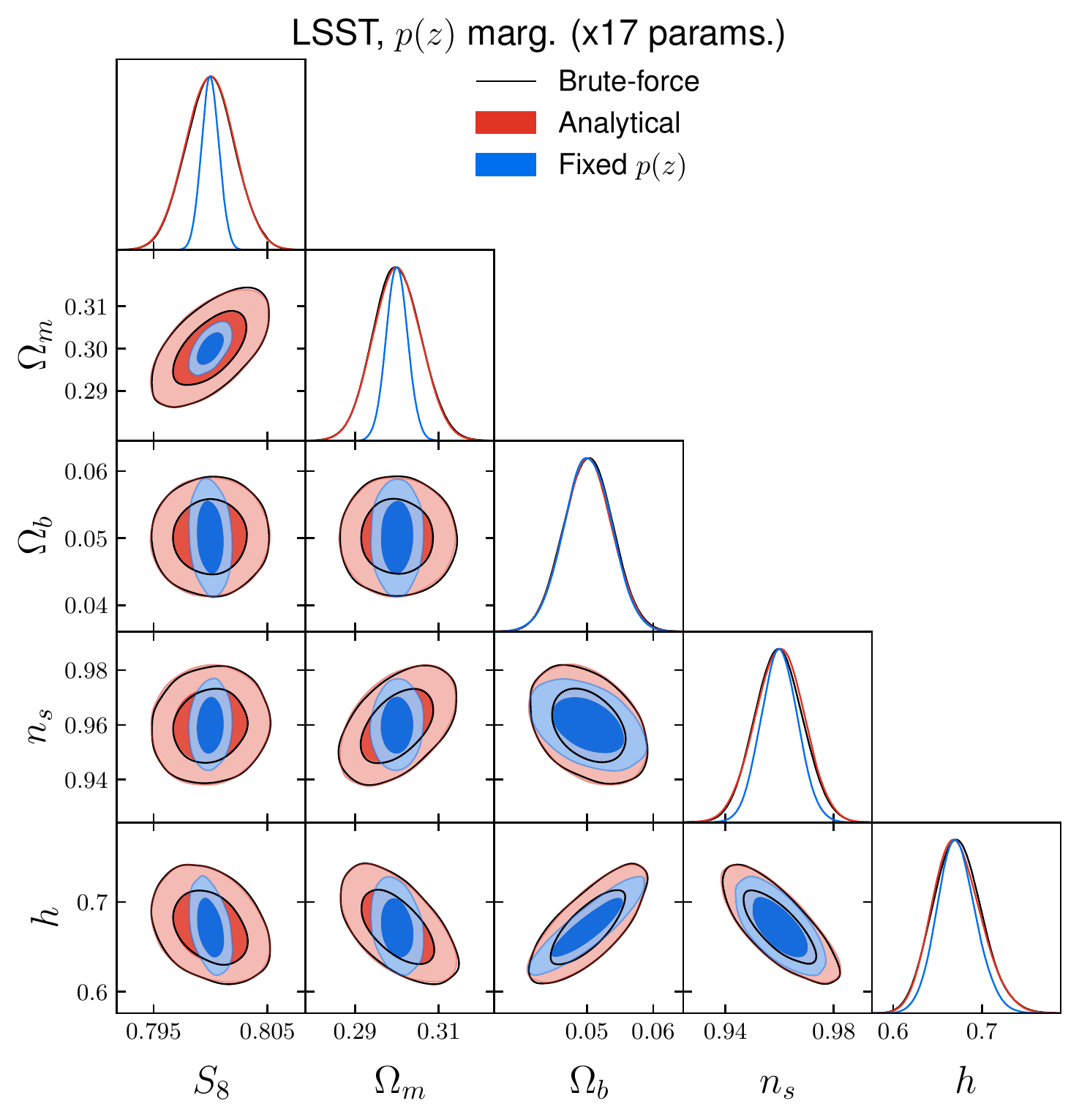}
      \caption{Same as Fig.~\ref{fig:DES_margdzwz}, but results are shown for a futuristic data vector with minimal noise meant to mimic the expected precision of LSST. The redshift distribution is less noisy compared with \desyo and is split into 6 tomographic bins. 
      We find that our analytic marginalisation recipe yields virtually equivalent results to the standard method for marginalising over redshift uncertainties by sampling directly the shift and width redshift parameters.}
      \label{fig:LSST_margdzwz}
    \end{figure*}
    
    \begin{figure}
      \centering
      \includegraphics[width=0.48\textwidth]{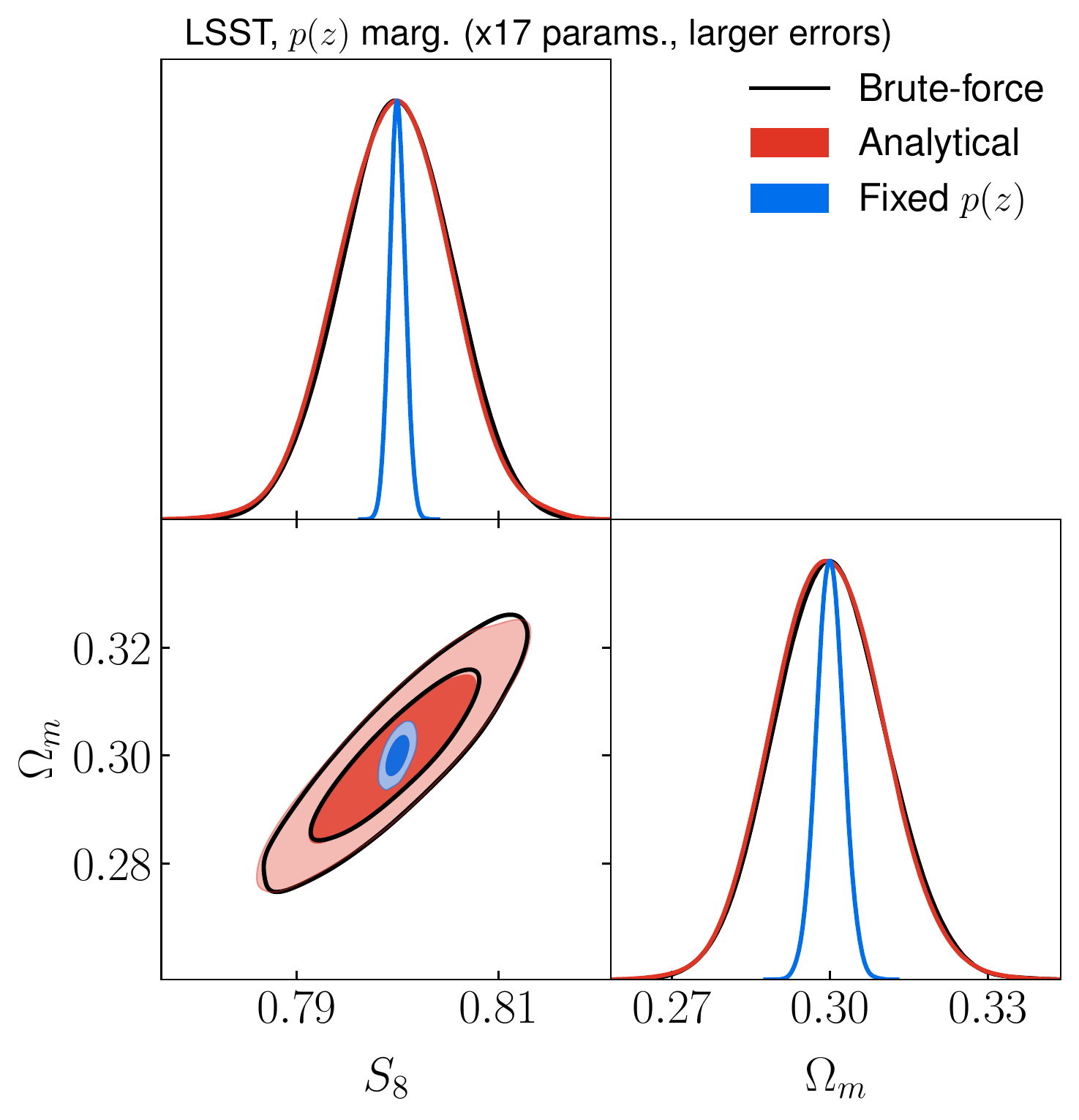}
      \caption{Same as Fig.~\ref{fig:LSST_margdzwz} except for the mean redshift distribution is the same as that of \desyo, but the errors on the nuisance parameters are quadrupled. As before, we find that our method of analytic marginalisation works well despite the larger errors on the $p(z)$ measurements. Thus, we assert that the approximations in Section~\ref{ssec:tomo.linear} hold even in the conservative scenario of large photometric uncertainties in futuristic data.}
      \label{fig:LSST_quad_margdzwz}
    \end{figure}
    We begin by focusing on calibratable systematics, for which we will follow the procedure described in Section \ref{ssec:tomo.linear}: we will linearise the dependence of the theoretical prediction with respect to these parameters, for which a relatively tight prior can be obtained, and simply modify the covariance matrix as in Eq.~\eqref{eq:modcov}, with ${\sf T}$ evaluated at a fixed set of parameters (fixing all cosmological parameters to the best-fit values found by {\sl Planck} \citep{2020A&A...641A...6P} and all bias parameters to their best-fit values). At this stage, we will thus only consider the nuisance parameters describing the uncertainty in the redshift distributions of the tracers under study, described in the right column of Table \ref{tab:priors_3x2pt}. All other parameters (cosmological, bias, and intrinsic alignment parameters) will for now be marginalised ``brute-force'' (i.e. treating them as free parameters in the MCMC chains). For simplicity, we will consider only linear bias, using scales $k<0.15\,{\rm Mpc}^{-1}$, as described in Section \ref{ssec:tomo.like}.

    We first compare the performance of the method when applied to the 3$\times$2pt analysis of \desyo data (Section~\ref{sssec:tomo.data.des}). The results are shown in Fig.~\ref{fig:DES_margdzwz} and summarised in Table~\ref{tab:marg_dzwz}, with the exact results shown as black contour lines, and the results of the analytical marginalisation shown in red. We find that using the analytic marginalisation technique not only yields contours that are almost indistinguishable from those obtained by the traditional approach, but also does so significantly faster (by a factor $\sim10$) with many fewer parameters. The blue contours in the figure show the constraints found by fixing the nuisance parameters to their prior means, instead of marginalising over them. In this case, we observe that the redshift distribution uncertainties cause only a mild broadening of the marginalised contours, which is not very challenging for the analytical approximate marginalisation to reproduce.

    In order to explore a more challenging scenario, we now move on to the case of an LSST-like 3$\times$2pt dataset, as defined in Section~\ref{sssec:tomo.data.lsst}. We will assume the same prior uncertainties used in the analysis of the \desyo data. This will allow us to quantify the validity of the analytical marginalisation approach in a conservative scenario, in which, in spite of the much higher sensitivity of Stage-IV data, the precision with which we are able to calibrate redshift distributions has not improved with respect to the performance achieved with current data. Furthermore, we will explore an additional case in which the prior uncertainties are 4 times larger than the \desyo ones. The reasoning behind this second test is two-fold: on one hand, our marginalisation method is an approximation that works in the regime where photometric uncertainties are linearisable and testing when that assumption breaks is essential; on the other hand, it is likely that, at the highest redshifts, and for the faintest samples, the $p(z)$ uncertainties will be somewhat larger for LSST than those of current surveys, especially at high redshifts. Hence, quadrupling the errors is an important worst-case scenario to consider.

    It is important to note that the results in this section are not meant to be interpreted as forecasts on the constraining power of LSST on cosmological parameters, but only as a quantification of our ability to analytically marginalise over photometric uncertainties when inferring the underlying cosmology. A more thorough analysis would include a realistic treatment of the LSST true redshift distribution and noise, and a more careful treatment of galaxy bias and intrinsic alignments. As such, the results presented here give us a conservative estimate of the effect of analytic marginalisation on cosmology constraints.

    We present the results of the first scenario (i.e., $p(z)$ errors matching DES) in Fig.~\ref{fig:LSST_margdzwz}. The marginalised constraints on all cosmological parameters are virtually unchanged when we switch from the brute-force to the analytical marginalisation. This latter method is therefore successful at recovering accurate marginalised constraints on cosmological parameters. The figure shows the constraints on all cosmological parameters, to highlight that the result extends even to the parameters that 3$\times$2pt datasets are less sensitive to: $\Omega_b$, $n_s$ and $h$. Also shown in blue are the constraints found assuming perfect knowledge of the redshift distributions (i.e. fixing all $p(z)$ parameters). In this case, the uncertainties on the redshift distributions have a much larger effects than for the \desyo data, inflating the uncertainties on $S_8$ by a factor $\sim2.6$. Thus, even though marginalising over the redshift distribution parameters has an outsized effect on the final constraints, the analytical approximation approach is able to capture it almost exactly.
    \begin{figure*}
      \centering
      \includegraphics[width=0.48\textwidth]{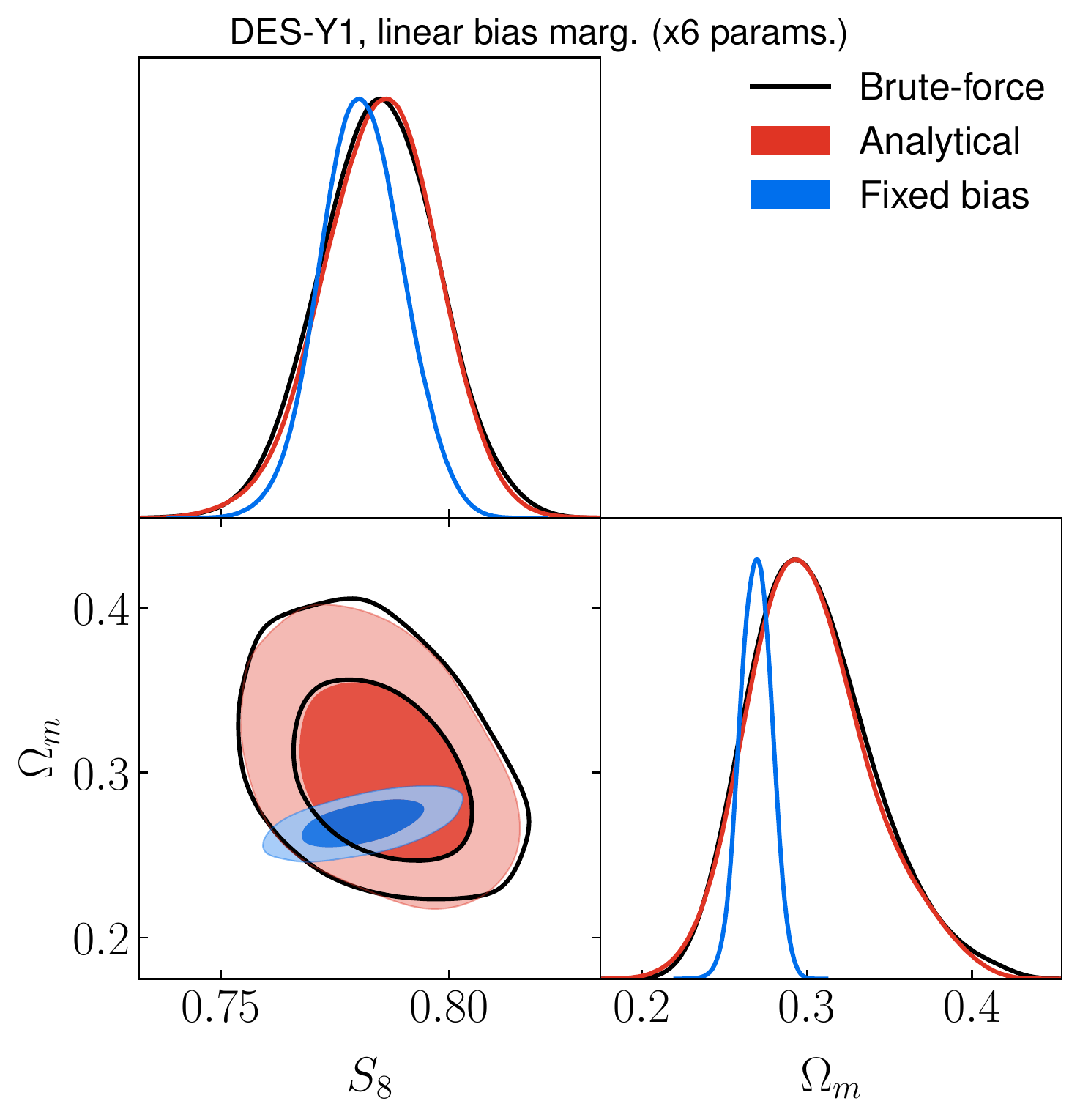}
      \includegraphics[width=0.48\textwidth]{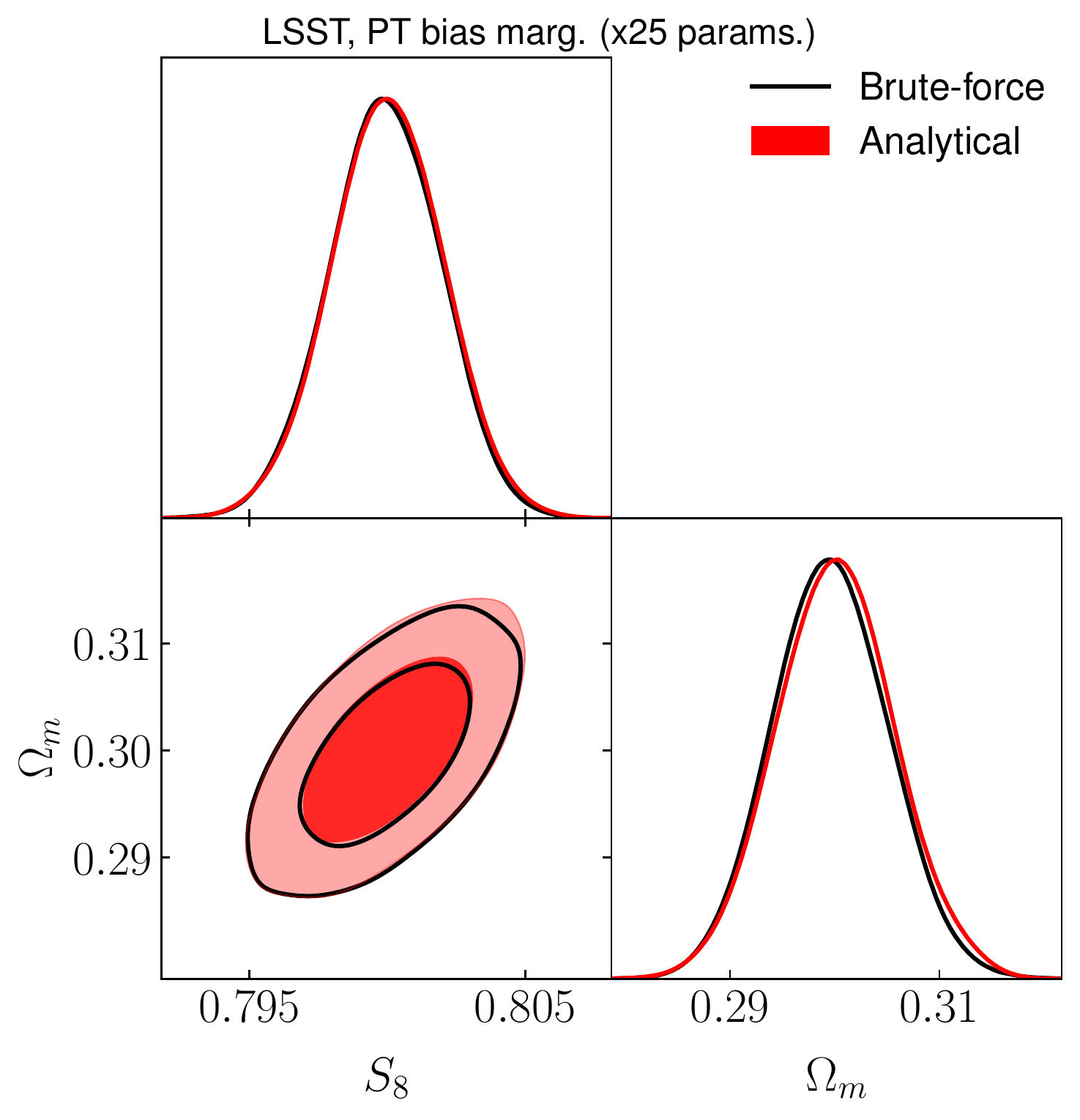}
      \caption{On the left, same as Fig.~\ref{fig:DES_margdzwz}, but here we marginalise over the bias parameters using the Laplace approximation. This reduces the parameter space to only the cosmological parameters without degrading the accuracy of the constraints. On the right, as Fig.~\ref{fig:LSST_margdzwz}, but here we marginalise over the perturbation theory bias parameters via the Laplace approximation. The agreement with the full numerical marginalisation (black) is almost perfect.}
      \label{fig:bias_tight}
    \end{figure*}

    Similarly reassuring is our result of quadrupling the uncertainty in the redshift nuisance parameters, shown in Fig.~\ref{fig:LSST_quad_margdzwz} in the $(\Omega_m,S_8)$ plane. In this case, the increased redshift distribution uncertainties broaden the final constraints on $S_8$ by up to a factor $\sim$10. In spite of this, we find that the analytic marginalisation method not only yields virtually the same constraints on the cosmological parameters, but does so 3-10 times faster than the traditional approach (see Table \ref{tab:marg_dzwz}). This implicitly validates the approximation that a first-order expansion of the theory data vector with respect to a change in redshift distribution is sufficient, even for prior uncertainties on $p(z)$ that are substantially worse than those achieved by current datasets.

  \subsection{Bias parameters: tight posteriors}\label{ssec:res.btight}

    Let us now focus on the bias parameters. In this case, the absence of an informative prior prevents us from linearising the dependence of the theory on these parameters, and we must therefore calculate the profile and Laplace terms numerically. Moreover, since the dependence of the $\chi^2$ on these parameters is not quadratic, marginalising over them may lead to significant volume effects (in the form of biases) in the marginalised posterior for the cosmological parameters. As a reminder, in the case of quadratic dependence of the $\chi^2$ on the parameters, the approximation would be exact, and our approximation would automatically take care of the volume effects. We will start by exploring two situations in which the data is sensitive enough to measure these bias parameters accurately, in which case, as we saw in Section \ref{sec:like}, the Laplace term and volume effects are small.
    
    In the first case, we make use of the \desyo data, marginalising only over a single linear galaxy bias parameter per clustering bin, as well as an intrinsic alignment amplitude (i.e. a total of 6 nuisance bias parameters). This roughly coincides with the analysis choices made for the official \desyo analysis. The results are shown in the left panel of Fig. \ref{fig:bias_tight}. The exact marginalised constrains (solid black contours) are accurately recovered by the Laplace approximation (red contours). While the former are obtained by running an MCMC with 11 free parameters (5 cosmological, 6 nuisance parameters), the latter involve only a 5-dimensional parameter space, which is therefore significantly simpler to explore. Concretely, the 5-parameter chain converged 3 times faster than the 11-parameter chain (see Table \ref{tab:marg_dzwz}). The blue contours in the same figure shows the constraints obtained after fixing the bias parameters to the best-fit values found by DES \citep{1708.01536}. Fixing the galaxy bias shifts the cosmological constraining power from the cosmic shear data to the higher signal-to-noise clustering data, thus significantly reducing the uncertainties. Note that, although the red contours show the result of the full Laplace approximation (i.e. profile $+$ Laplace contributions), the Laplace contribution is negligible, and the profile term is enough to recover the marginalised posterior. 

    In order to explore the performance of the method with a significantly larger number of nuisance parameters, while still remaining in the regime where these parameters can be well constrained by the data, we now move to the LSST-like synthetic dataset, making use of the second-order perturbative expansion of Eq.~\eqref{eq:bias_pt} to describe galaxy bias. In this case, we include 4 free bias parameter in each of the 6 clustering redshift bins, adding up to a total of 25 nuisance parameters when combined with the intrinsic alignment amplitude. The results are shown in the right panel of Fig. \ref{fig:bias_tight}, again as black lines for the brute-force marginalisation, i.e. considering all 30 free parameters, and as red contours for the 5-parameter Laplace approximation. As before, we find that the approximation is able to recover the marginalised constraints almost exactly. The impact of volume effects is also heavily suppressed, shifting the marginalised contours by less than $0.3\sigma$ (not shown in the figure).

    \begin{figure*}
      \centering
      \includegraphics[width=0.48\textwidth]{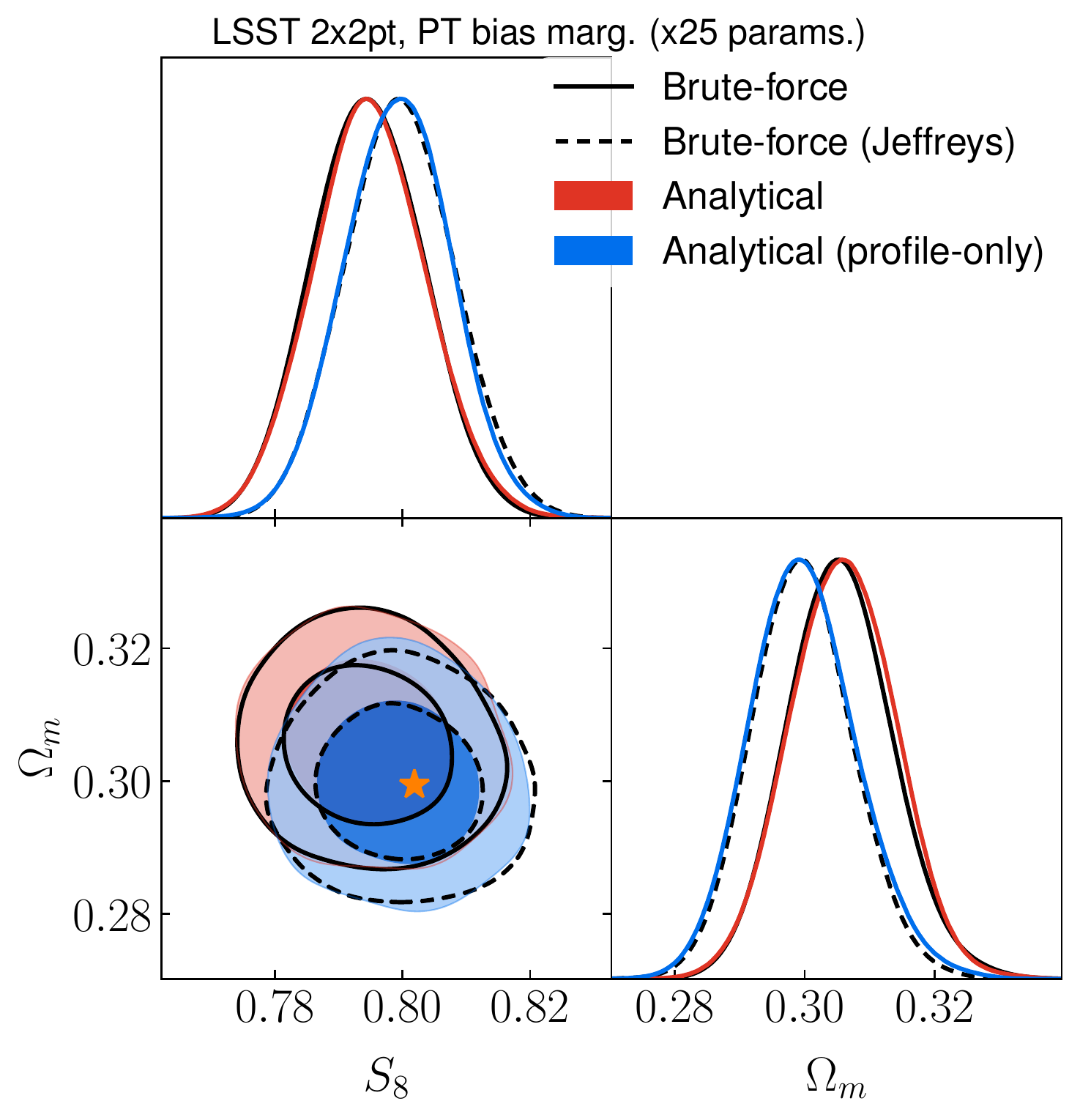}
      \includegraphics[width=0.48\textwidth]{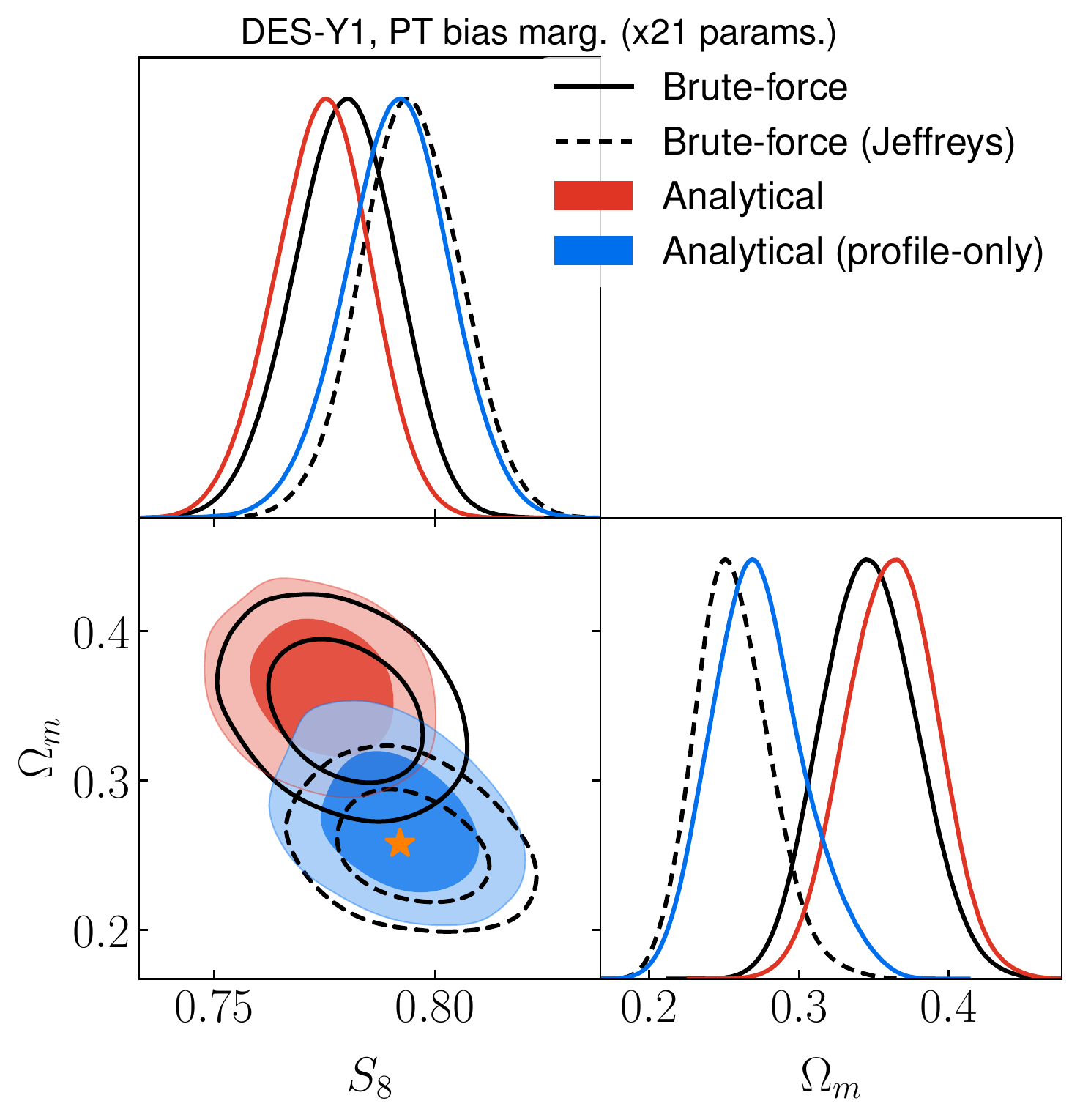}
      \caption{Demonstration of volume effects in the case of using less constraining data. In order to recover the best-fit parameter values (denoted with a star), when applying brute-force marginalisation, one needs to make use of Jeffreys prior (compare solid with dashed black line). Similarly, when analytically marginalising over the PT biases, one needs to remove the Laplace term to avoid biasing the constraints (compare blue and red contours). The plot on the left shows the constraints for a 2x2pt analysis with a LSST-like dataset, while on the right we show constraints using the DES-Y1 data. In the latter case, the volume effects become more pronounced, as the data has less constraining power.}
      \label{fig:lsst_2x2pt}
    \end{figure*}
    
    Note that, in both of these cases, besides the bias parameters, we have also marginalised over a number of redshift distribution uncertainty parameters (14 for \desyo, 17 for LSST), thus reducing the model dimensionality from 47 and 28 for LSST and \desyo, respectively, to only 5 cosmological parameters. Obtaining converged MCMC chains for these 5 cosmological parameters takes approximately 2 hours for the LSST-like dataset, a factor $\sim6$ faster than the chains with only the $p(z)$ parameters analytically marginalised over, and a factor $\sim 30$ times faster than the full brute-force chains. The magnitude of these speed gains, however impressive, must be taken with a pinch of salt. The performance of MCMC sampling may depend significantly upon the design of the likelihood code, and whether it allows the sampler to decompose the space between ``fast'' and ``slow'' parameters, over-sampling the former, and making use of ``dragging'' techniques \citep{2013PhRvD..87j3529L,2005math......2099N}. The fast-slow split allows one to effectively marginalise over the fast subspace, and becomes particularly powerful in the presence of a large number of nuisance parameters on which the likelihood has a computationally simple dependence (as is the case of the bias parameters in our model). To provide a fairer assessment of the computational gains obtained using the Laplace approximation, we wrote an optimised version of the 3$\times$2-point likelihood that allows {\tt cobaya} to exploit the fast nature of the bias parameters as efficiently as possible (assuming all $p(z)$ parameters are fixed). In this case, the speed-up factor for the cases explored here ranged between $\sim1.5$ and $\sim11$, with the highest performance improvement achieved for the LSST 2$\times$2pt data discussed in the next section. Ultimately the performance difference depends on the design of the likelihood code, the complexity in the parameter dependence of the likelihood, and the efficiency of the $\chi^2$ minimisation method used to calculate $\badpar_*$. On this latter point, we find that the Gauss-Newton method used here typically achieves convergence after only 10-20 iterations. Thus finding the best-fit bias parameters is significantly faster than calculating the cosmology-dependent power spectrum templates of Eq.~\eqref{eq:cl_bias_model}.

  \subsection{Bias parameters: loose posteriors and volume effects}\label{ssec:res.bloose}
    In the cases explored in the previous sections, the data was able to constrain the nuisance parameters sufficiently well, and all volume effects associated with the choice of parametrisation were subdominant. Let us now study the ability of the Laplace approximation to capture these volume effects when they become more prominent.
    
    First, let us consider the case of a ``2$\times$2pt'' combination of the LSST data, consisting only of the galaxy-galaxy and galaxy-shear power spectra, excluding the shear-shear auto-correlation, and analysed under the perturbative bias expansion of Eq.~\eqref{eq:bias_pt}. The logic for focusing on this scenario is that, in this case, the model relies completely on galaxy clustering to constrain cosmological parameters, and is therefore more sensitive to the complexity of the galaxy bias parametrisation (which dominates the dimensionality of the nuisance parameter space). The resulting constraints on $\Omega_m$ and $S_8$ are shown in the left panel of Fig. \ref{fig:lsst_2x2pt}. The solid and dashed black contours show the results using brute-force marginalisation, assuming no prior and a Jeffreys prior respectively. The red contours then show the same constraints using the Laplace approximation (including both profile and Laplace terms), while the blue contour shows the result of including only the profile likelihood contribution. The position of the best-fit parameters is indicated by the orange star. We find that the profile-only approximation is able to recover well the exact marginalised constraints assuming a Jeffreys prior, and the full Laplace approximation recovers the constraints found in the absence of this prior. In this case we see a clear $\sim0.8\sigma$ shift in the cosmological constraints due to volume effects associated with the large number of bias parameters, although the Laplace term is able to capture this with high accuracy.

    As a second example, we explore the possibility of using the second-order perturbative bias expansion of Eq.~\eqref{eq:bias_pt} to analyse the \desyo data. The lower sensitivity of this data compared to LSST should reduce its ability to constrain the higher-order bias parameters, and increase the impact of volume effects. The results are shown in the right panel of Fig. \ref{fig:lsst_2x2pt}, using the same color scheme as the last figure. In this case we can see that volume effects are a lot more significant, and can shift the confidence contours of the cosmological parameters by almost $3\sigma$ with respect to their global best-fit value, marked by the orange star. We can also see that, in this more extreme case, the Laplace approximation starts to fail, manifesting mild shifts of $\sim0.3-0.5\sigma$ with respect to the exact marginalised constraints. This provides us with a useful ``rule of thumb'' to determine the reliability of the Laplace approximation: if a significant ($>1\sigma$) shift is found between the marginalised contours obtained using the profile likelihood and those obtained accounting also for the Laplace term, the approximation may start to fail, and a brute-force marginalisation over the nuisance parameters using a Jeffreys prior is needed to obtain accurate results.   This test can be done without running the MCMC twice: one can run MCMC chain using Laplace approximation while also saving the profile likelihood values at each MCMC step and employing importance sampling to derive the profile likelihood posteriors. Nevertheless, even in these cases, we find that the Laplace approximation is able to recover the region of parameter space preferred by the data reasonably well, and can thus be used as a fast way to characterise this region that can then be refined (e.g. via importance sampling). Another good criterion for testing the accuracy of the Laplace approximation in the case of linear parameters is to explore the difference between the Laplace approximation treatment, computed with $\mathcal{F}$, and the Jeffreys prior treatment, computed with $F$ (defined in Eq.~\ref{eq:fisher}), since this difference should be non-zero only when the parameters are non-linear and would otherwise demarcate the breakdown of our approximation.

    In an extreme case, such as the one we just discussed, it is worth asking which of the constraints in Fig. \ref{fig:lsst_2x2pt} are the ``correct'' ones. The presence of such large volume effects, combined with the arbitrariness of using flat priors for the nuisance parameters, would imply that the use of a Jeffreys prior should produce the most correct results. However, as we noted in Section \ref{ssec:like.jeffrey}, the Jeffreys prior is not guaranteed to cancel out all volume effects. One might thus argue that the profile of the full likelihood, since it is by construction centered on the best-fit parameters, could provide an equally reasonable representation of the favoured confidence region. One might, however, view this choice critically, as it neglects the volume effects associated with the physical (in our case, cosmological) parameters. Finally, we could adopt a completely frequentist approach, determining the confidence intervals of all parameters by extracting the maximum-likelihood estimate for the data and for a suite of simulated datasets compatible with it.

\section{Conclusions}\label{sec:conc}
  Forecasts of the next decade in cosmology predict that meaningful constraints on fundamental unknowns such as the mass of neutrinos and the nature of dark matter and dark energy will come from multi-scale, multi-tracer efforts, encompassing a wide range of probes and redshifts. It is for this reason that the efficient analysis of joint data sets combining low- and high-redshift probes in an accurate manner is of great importance to the field of large-scale structure analysis. However, this comes at the cost of adding a colossal number of nuisance parameters characterising the observational and theoretical systematic uncertainties of all probes involved, which can noticeably slow down the sampling of the parameter space. In galaxy clustering and weak lensing joint studies, the most significant obstacles to overcome are the accurate modeling of the redshift distribution, the galaxy bias relation, and intrinsic alignments. Not doing so correctly can bias the inferred cosmological parameters, but also accounting for these effects via more elaborate models is a major inhibitor of efficient sampling.

  In this paper, we introduce a formal approach for speeding up the sampling process in the presence of a large number of nuisance parameters, and investigate the accuracy of our method when applying it to photometric survey data. In particular, we study the current \desyo data set as well as a synthetic data vector from an LSST-like survey to validate whether the fast analytic method proposed in this work is capable of reproducing the posterior contours and constraints one arrives at when adopting the traditional method of diligently varying tens of nuisance parameters.

  In Section~\ref{sec:like}, we describe the general methodology behind analytically marginalising over any nuisance parameter by approximating the marginal distribution, Eq.~\eqref{eq:laplace}, as consisting of a ``profile'' term, centered on the best-fit nuisance parameters, and a ``Laplace'' term, associated with the quadratic contribution. We then consider the special case of a Gaussian likelihood, showing that the Laplace term can be associated with volume effects, which the profile term alone is free from, being approximately equivalent to imposing a Jeffreys prior. This argument becomes exact when studying the case of nuisance parameters that contribute linearly to the theory vector. In Section~\ref{sec:tomo}, we introduce the cosmological analysis of photometric survey analysis and the specific problems associated with it. In particular, we introduce the relevant summary statistics utilised when performing ``3$\times$2pt'' analysis and the two dominant sets of nuisance parameters associated with that analysis: redshift distribution ($p(z)$) calibration parameters and bias/IA model parameters. We then provide details of how our model enables a fast marginalisation over these and how it can be applied to current and future data sets.

  We summarise our results in Section~\ref{sec:res}. The first half of that section deals with the $p(z)$ parameters. Since these parameters have external priors, the dependence of the theory vector on them can be linearised around the center of that prior, and the Laplace term can be ignored. Their impact can then be incorporated in a pre-sampling step by simply modifying the covariance matrix (see Eq.~\eqref{eq:modcov}). We showed (Figs.~\ref{fig:DES_margdzwz} and \ref{fig:LSST_margdzwz}), that this method is able to obtain constraints on cosmological parameters that are indistinguishable from those found via brute-force marginalisation, while performing 5-10 times better in terms of computational time, even when considering calibration priors that are significantly worse than those that can be achieved with current data.

  We next focused on the marginalisation over bias parameters, using the full Laplace approximation, finding the maximum of the conditional posterior distribution on the fly. We find that, when the data is able to place significant constraints on all model parameters, the Laplace approximation is an excellent representation of the marginal distribution, and that the Laplace term becomes subdominant with respect to the profile likelihood. In the presence of less constraining data, volume effects associated with the choice of nuisance parametrisation become relevant and, when mild, can be accurately captured by the Laplace term. In these cases, we also find that the profile likelihood is almost indistinguishable from the marginal distribution when adopting a Jeffreys prior, cancelling these volume effects. When volume effects become more relevant, significantly shifting the parameter region favoured by the marginal distribution (e.g. by more than $\sim1\sigma$), the Laplace approximation begins to fail, although the position and extent of the favoured region of parameter space are still qualitatively well described by it (both with or without a Jeffreys prior). Thus, even in this case, the method can be used to identify this region and then refine it with a standard MCMC.

  Through this paper we have assumed that the likelihood in the nuisance parameters is unimodal. Obviously, for multimodal likelihoods with comparable posterior volumes in each mode, the approximations we use will fail. Since the likelihood is in general quartic in bias parameters, the likelihood can in principle have up to three local maxima in each bias direction making multimodal likelihoods a possibility. This has been observed ``in the wild'' in fits to galaxy auto-power spectrum using higher order bias parameters \citep[e.g.,][]{2022PhRvD.105l3518G}, but typically disappears with sufficiently aggressive scale cuts and when employing the full 3$\times$2pt data vector. These problems can be detected by initialising the Newton-Raphson optimiser at different starting points and noting existence of multiple maxima. 

  The elimination of nuisance parameters is a nontrivial task. Caution must be taken when departing from our assumption of Gaussian data with a model-independent covariance matrix. While in this specific application, we find the profile likelihood to give unbiased parameter constraints and the marginal distribution to be biased, this does not hold in general. Common tasks in data analysis (such as the estimation of the variance from Gaussian data) have an opposite outcome, with the marginal distribution peaking at the unbiased parameter estimate, while in other cases neither method seems to perform optimally \citep{10.2307/2676644}. It is therefore crucial to consider the role of volume effects in every specific likelihood model.
  
  The reduction in the dimensionality of the parameter space afforded by the Laplace approximation is accompanied by a boost in the speed with which convergence in the associated MCMC chains is achieved, and we find improvements by a factor $\sim2-15$. Achieving this performance improvement is greatly aided by the simplicity with which the theory prediction depends on the bias and IA nuisance parameters, which allows us to write down its derivatives analytically (see Section \ref{ssec:tomo.bias}), and to evaluate them quickly (in $O(10^{-3}\,{\rm s})$). There may be more sophisticated astrophysical models that do not conform to this simple structure (e.g. physically motivated models for the redshift dependence of bias terms, halo-based models, etc.), and for which computing derivatives analytically is impractical or unfeasible. A more general application of this method in these situations would thus require the use of automatic differentiation techniques \citep{2018arXiv181105031M}, able to efficiently calculate these gradients regardless of how the model depends on the nuisance parameters. The Laplace approximation is by far not the only application that benefits from having efficient access to derivatives of the likelihood with respect to some parameters. More general sampling methods including HMC \citep{HMC} or variational inference \citep{2016arXiv160100670B} often require the use of likelihood gradients, as do efficient minimisation methods, or the calculation of Jeffreys priors. Additionally, although here we have focused on the volume effects associated with the marginalisation over nuisance parameters, the non-linear way in which all cosmological parameters enter the likelihood implies that degeneracies between them can also give rise to biasing volume effects within the subspace of cosmological parameters. A Jeffreys prior (or any other form of correction for volume effects) should therefore be used routinely in cosmological parameter inference, although this is normally prevented by the need to estimate derivatives efficiently. The development of fully automatically-differentiable cosmological theory predictions is therefore of paramount importance in the context of current and future experiments.

  In conclusion, our method produces satisfactory results in virtually all of the explored regimes. Thanks to the huge gain in sampling efficiency this approach offers, it can be used to enable the joint analysis of multiple galaxy surveys, which is typically hindered by the gigantic number of nuisance parameters that need to be sampled. Fast marginalisation methods such as the one proposed in this work can help us address some of the inconsistencies in the present analysis of cosmological data. For example, current constraints find that both the amplitude of clustering ($S_8$) and the energy density of matter ($\Omega_m$) inferred from the low-redshift analysis of galaxy and weak lensing surveys are systemically lower than those obtained from high-redshift probes such as the CMB \citep[for a review, see][]{2022NewAR..9501659P}. In the near term, we plan to adopt the method proposed in this paper to speed up the sampling process and analyze jointly the currently (publicly) available photometric survey data in an effort to tackle some of the obstacles standing in the way of gaining a fundamental understanding of our Universe.

\begin{acknowledgments}
  We would like to thank Deaglan Bartlett, Harry Desmond, Simone Ferraro, Arrykrishna Mootoovaloo, and Martin White for useful discussions. BH is supported by the Miller Institute for Basic Research in Science and the Chamberlain Fellowship and Lawrence Berkeley National Lab. KW is funded by a SISSA Ph.D. fellowship. SA is funded by a Kavli/IPMU doctoral studentship. DA is supported by the Science and Technology Facilities Council through an Ernest Rutherford Fellowship, grant reference ST/P004474. CGG is supported by European Research Council Grant No:  693024 and the Beecroft Trust. JRZ is supported by an STFC doctoral studentship.

\end{acknowledgments}

\bibliographystyle{mnras}
\bibliography{main}

\appendix

\section{Summary of results}
\begin{table*}
  \begin{center}
  \begin{tabular}{c | c c c c c c}
     \hline\hline
     Method & $S_8$ & $\Omega_m$ & $\Omega_b$ & $n_s$ & $h$ & Converge time \\ [0.5ex]
     \hline
     DES 3$\times$2pt: brute-force & $0.784\pm 0.017$ & $0.304^{+0.029}_{-0.041}$ & -- & -- & -- & 338.34 Hrs \\ [1ex]
    DES 3$\times$2pt: analytic & $0.783\pm 0.017$ & $0.305^{+0.030}_{-0.042}$ & -- & -- & -- & 31.74 Hrs \\ [1ex]
    DES 3$\times$2pt: fixed $z$ & $0.780\pm 0.015$ & $0.306^{+0.029}_{-0.040}$ & -- & -- & -- & 30.22 Hrs \\ [1ex]
     LSST 3$\times$2pt: brute-force & $0.7999\pm 0.0021$ & $0.3001\pm 0.0057$ & $0.0501\pm 0.0036$ & $0.9600\pm 0.0090$ & $0.672^{+0.024}_{-0.029}$ & 58.24 Hrs \\ [1ex]
    LSST 3$\times$2pt: analytic & $0.8000\pm 0.0021$ & $0.3001\pm 0.0057$ & $0.0502\pm 0.0037$ & $0.9597\pm 0.0089$ & $0.673^{+0.026}_{-0.028}$ & 13.00 Hrs \\ [1ex]
    LSST 3$\times$2pt (4x): brute-force & $0.7996\pm 0.0056$ & $0.300\pm 0.010$ & $0.0501\pm 0.0038$ & $0.9599\pm 0.0097$ & $0.673^{+0.028}_{-0.035}$ & 23.19 Hrs \\ [1ex]
    LSST 3$\times$2pt (4x): analytic & $0.7999\pm 0.0054$ & $0.300\pm 0.010$ & $0.0503\pm 0.0037$ & $0.9598\pm 0.0099$ & $0.673^{+0.029}_{-0.035}$ & 8.36 Hrs \\ [1ex]
    LSST 3$\times$2pt: fixed $z$ & $0.79998\pm 0.00078$ & $0.3000\pm 0.0026$ & $0.0501\pm 0.0036$ & $0.9599\pm 0.0068$ & $0.672^{+0.020}_{-0.023}$ & 12.39 Hrs \\ [1ex]
    \hline
    \hline
  \end{tabular}
  \end{center}
  \caption{Constraints on the cosmological parameters from our 3$\times$2pt analysis applied to a current (\desyo) and future (LSST-like) data set, comparing the performance of two marginalisation approaches: analytic and brute-force (described in Section~\ref{ssec:tomo.linear}). ``Analytic'' refers to the new method proposed in this work, in which one accounts for photometric uncertainties prior to sampling the parameter space, whereas ``brute-force'' refers to the standard method of introducing $\sim$10 ``shift'' and ``width'' redshift calibration parameters (see Table~\ref{tab:priors_3x2pt}). ``Fixed $z$'' considers the unrealistic scenario of completely ignoring photometric uncertainties.}\label{tab:marg_dzwz}
\end{table*} 
\begin{table*}
  \begin{center}
  \begin{tabular}{c | c c c c c c}
    \hline\hline
    Method & $S_8$ & $\Omega_m$ & $\Omega_b$ & $n_s$ & $h$ & Converge time \\ [0.5ex]
    \hline
    DES-Y1 linear bias: brute-force & $0.785\pm 0.013$ & $0.303^{+0.030}_{-0.043}$ & -- & -- & -- & 7.1 Hrs \\ [1ex]
    DES-Y1 linear bias: analytic & $0.785\pm 0.012$ & $0.302^{+0.030}_{-0.041}$ & -- & -- & -- & 2.5 Hrs\\ [1ex]
    DES-Y1 linear bias: fixed $b$ & $0.7810\pm 0.0090$ & $0.2691\pm 0.0094$ & -- & -- & -- & 2.1 Hrs \\ [1ex]
    DES-Y1 PT bias: brute-force, no J.P. & $0.780\pm 0.012$ & $0.347\pm 0.031$ & -- & -- & -- & 16.5 Hrs\\ [1ex]
    DES-Y1 PT bias: analytic, full Laplace & $0.775\pm 0.011$ & $0.362\pm 0.030$ & -- & -- & -- & 6.5 Hrs\\ [1ex]
    DES-Y1 PT bias: brute-force, with J.P. & $0.795\pm 0.011$ & $0.256^{+0.022}_{-0.027}$ & -- & -- & -- & 45.5 Hrs\\ [1ex]
    DES-Y1 PT bias: analytic, profile only & $0.792\pm 0.012$ & $0.273^{+0.027}_{-0.034}$ & -- & -- & -- & 8.1 Hrs\\ [1ex]
    LSST 3$\times$2pt: brute-force & $0.7999\pm 0.0020$ & $0.2996\pm 0.0055$ & $0.0505\pm 0.0036$ & $0.9580\pm 0.0089$ & $0.676^{+0.025}_{-0.030}$ & 5.1 Hrs\\ [1ex] 
    LSST 3$\times$2pt: analytic & $0.8000\pm 0.0020$ & $0.3001\pm 0.0057$ & $0.0501\pm 0.0037$ & $0.9595\pm 0.0089$ & $0.672^{+0.025}_{-0.030}$ & 2.2 Hrs\\ [1ex] 
    LSST 2$\times$2pt: brute-force, no J.P. & $0.7946\pm 0.0086$ & $0.3055\pm 0.0081$ & $0.0482\pm 0.0044$ & $0.955\pm 0.037$ & $0.655^{+0.035}_{-0.044}$ & 35.1 Hrs\\ [1ex]
    LSST 2$\times$2pt: analytic, full Laplace & $0.7949\pm 0.0088$ & $0.3059\pm 0.0081$ & $0.0483\pm 0.0043$ & $0.955\pm 0.038$ & $0.654^{+0.037}_{-0.042}$ & 2.1 Hrs \\ [1ex]
    LSST 2$\times$2pt: brute-force, with J.P. & $0.7997\pm 0.0086$ & $0.2997^{+0.0072}_{-0.0081}$ & $0.0504\pm 0.0044$ & $0.954\pm 0.039$ & $0.680^{+0.040}_{-0.049}$ & 17.2 Hrs\\ [1ex]
    LSST 2$\times$2pt: analytic, profile only & $0.7993\pm 0.0083$ & $0.2998^{+0.0075}_{-0.0087}$ & $0.0506\pm 0.0043$ & $0.953\pm 0.040$ & $0.682^{+0.041}_{-0.049}$ & 6.4 Hrs\\ [1ex]
    \hline
    \hline
  \end{tabular}
  \end{center}
  \caption{As in Table~\ref{tab:marg_dzwz}, but here we show the results from our bias marginalization approach (described in Section~\ref{ssec:tomo.bias}). ``Analytic'' refers to the new method proposed in this work, in which one analytically incorporates the uncertainty associated with the bias parameters on-the-fly by adopting the Laplace approximation, whereas ``brute-force'' refers to the standard method of directly sampling and marginalizing over the bias parameters (see Table~\ref{tab:priors_3x2pt}). ``Fixed $b$'' considers the unrealistic scenario of setting the bias parameters to a constant value.}\label{tab:marg_bias} 
\end{table*}
  Table  \ref{tab:marg_dzwz} lists the constraints on cosmological parameters, and the time needed for the corresponding MCMC chains to converge for the different cases explored in Section \ref{ssec:res.dz}, where we marginalise over calibratable redshift distribution systematics using the linearisation technique described in Section \ref{ssec:tomo.linear}. Table \ref{tab:marg_bias} shows the same information for the cases explored in Sections \ref{ssec:res.btight} and \ref{ssec:res.bloose}, in which we make use of the full Laplace approximation to marginalise over galaxy bias and intrinsic alignment parameters.
\end{document}